\documentclass[
  reprint,
  superscriptaddress,
  nofootinbib,
  floatfix,
  amsmath,amssymb,
  aps, pra,
  colorlinks=true,
  citecolor=blue,
  linkcolor=blue,
  urlcolor=blue
]{revtex4-2}

\usepackage{graphicx}   
\usepackage{dcolumn}    
\usepackage{bm}         

\usepackage[table,dvipsnames]{xcolor} 
\usepackage{booktabs}
\usepackage{multirow}

\usepackage{siunitx}
\usepackage{placeins}   

\usepackage{hyperref}
\usepackage{orcidlink}

\newcommand{\AffCagliari}{University of Cagliari, Via Is Mirrionis 1, Cagliari, 09123, Italy}

\newcommand{\AffUNLP}{Instituto de F\'isica (IFLP-CCT-CONICET), Universidad Nacional de La Plata, C.C. 727, La Plata, 1900, Argentina}

\newcommand{\AffTUM}{Institute for Advanced Study, Technische Universit\"at M\"unchen, M\"unchen, Germany}



\begin{document}

\preprint{APS/123-QED}

\title{Distribution of Non-Locality On Quantum Random Circuits}


\author{Andrés Camilo Granda Arango\orcidlink{0000-0002-5255-2507}}
\affiliation{\AffCagliari}
\author{Federico Hernán Holik\orcidlink{0000-0002-6776-5281}}
\email[Contact author: ]{holik@fisica.unlp.edu.ar}

\affiliation{\AffCagliari}  
\affiliation{\AffUNLP}      
\author{Roberto Giuntini\orcidlink{0000-0003-2078-0297}}
\affiliation{\AffCagliari}
\affiliation{\AffTUM}
\author{Hector Freytes}
\affiliation{\AffCagliari}
\author{Giuseppe Sergioli\orcidlink{0000-0002-3650-5858}}
\affiliation{\AffCagliari}

~

\begin{abstract}
In this work we explore how different types of resources are distributed among the states generated by quantum random circuits (QRC). We focus on multipartite non-locality, but we also analyze quantum correlations by appealing to different entanglement and non-classicality measures. We analyze the violation of Mermin and Svetlichny inequalities in order to get a glance at the distribution of nonlocality and genuine multipartite nonlocality. Next, we compare universal vs non-universal sets of gates, to gain insight into the problem of explaining quantum advantage. By comparing the results obtained with ideal (noiseless) vs  noisy intermediate-scale quantum (NISQ) devices, we lay the basis of a certification protocol, which aims to quantify how robust is the resources distribution among the states that a given device can generate. We have implemented our non-locality-based benchmark on actual quantum processors with different architectures, in order to assess up to which point they are capable of reproducing the ideal results.
\end{abstract}

\maketitle



\section{\label{sec:level1}Introduction}

Quantum computers offer promising computational advantages for the years to come. There exist many operating devices nowadays that can be easily accessed using the cloud quantum computing model. This rapid growth is accompanied by claims of quantum advantage with near-term devices~\cite{Zuchongzhi2025,Willow2025,Madsen_Xanadu_2022}. Though the interpretation of quantum supremacy claims generates a heated debate and should be taken cautiously (see for example~\cite{Arute_2019_Sycamore} and~\cite{Liu_Sycamore_paper_is_a_fraud}), it is clear that the road map is promising. There are great chances that, in the near future, quantum computers will be able to perform tasks that are very hard for classical supercomputers, even if the path to fault-tolerant universal quantum computing and commercial applications remains challenging. 

Nowadays prototypes of quantum computers are noisy and present a variable performance among the different architectures. It is then relevant to characterize how capable these devices are of producing different types of resources that are considered nonclassical and necessary for computational advantage. In this work we focus on the capability of simulating nonclassical correlations using quantum random circuits. Our study shows how rich is the quantum state space a given quantum device can produce. We also analyze the differences of devices using only Clifford vs Clifford $+$ \texttt{T} sets of native gates. As is well known, the Clifford $+$ \texttt{T} set can be used to approximate any conceivable quantum circuit, given that it is a universal set. On the contrary, a QPU using only Clifford gates is not universal, and it is known that its outputs can be efficiently simulated by a classical probabilistic Turing machine~\cite{Cuffaro_Gottesman_Knill_Theorem}. Out of our comparison, we obtain valuable information about how resources are distributed in both cases, yielding a novel insight on discussions about quantum advantage.

There are many hypotheses to explain why quantum computers seem to have a speed up with regard to their classical cousins. Overall, the reasons are not completely clear. Recent studies point to different quantum resources, such as entanglement and non-locality~\cite{jozsa1997entanglement}, quantum coherence~\cite{Matera_Coherence_as_resource} and contextuality~\cite{ContextualityNature-2011}. But none of these quantum features seems to suffice on its own to explain quantum advantage, due to results such as the Gottesman-Knill theorem~\cite{Gottesman_1998} (though see for example the discussion in~\cite{Cuffaro_Gottesman_Knill_Theorem}). Furthermore, calibration and benchmarking quantum computers is usually a complicated business. As extant quantum processors are very sensitive to noise, it is important to develop different tools to assess their capabilities (see, for example,~\cite{wootton2018benchmarking, QuantumVolume_Photonic, Emerson_2005, Emerson_2019}). The general question in which we are interested is: how are quantum resources distributed among the states that a concrete quantum device can produce? In this work we propose a method to assess that. 

We study the resources produced in different architectures of quantum computers using states generated by quantum random circuits (QRC)~\cite{Bouland2019}. We focus on non-locality by using the Mermin~\cite{Mermin_1990} and the Svetlichny inequalities~\cite{Svetlichny_PRD,Seevinck_Svetlichny_PRL}. For completeness, we compare non-locality with entanglement by appealing to different measures, such as tangle~\cite{Coffman_2000}
and entanglement for multi-partite systems~\cite{MeyerWallach2002}. We also analyze a quantifier of how much non-Clifford gates are present in a given circuit which is known as quantum magic~\cite{Oliviero2022}. 

We first perform a theoretical study based on numerical simulations using the Amazon Braket SDK~\cite{braket}. We analyze the performance of the devices with different levels of noise and number of shots, and compare universal (Clifford $+$ \texttt{T}) vs non-universal (Clifford) sets of elementary gates. Given a set of physical features which are thought of as a resource for quantum computing (e.g., non-locality), we quantify up to which extent a device can give place to a state space rich enough to have the desired resources. The most salient difference between the universal and non-universal sets of gates analyzed in this work, is the discreteness in the possible values of the resources distribution of the latter. Indeed, while a device using only the Clifford set of gates can produce states with maximal entanglement and non-locality, the possible values of these magnitudes are highly concentrated in a few discrete values. We also show how the resources distribution is deteriorated with noise for architectures based on universal sets of gates. 
  
Our findings are also relevant for the certification of quantum devices. By quantifying to which extent the resources distribution departs from the ideal case, one can have a global evaluation of the effective state space that a concrete quantum processing unit (QPU) can generate. The proposed methodology is different from a test merely based on preparing a concrete family of states, such as generalizations of the Greenberger--Horne--Zeilinger (GHZ) state, or testing a particular algorithm. Our developments allow to assess the capabilities of a quantum device as a whole, by using one of the main features of fundamental physics, namely, non-locality (see also~\cite{Cabello-Non-locality-benchmarking,Baumer2021,Lanyon2014} for other studies based on nonclassical correlations in relation to quantum processors). Our methodology can also be used to test the performance of a specific set of target qubits, a task that is relevant for calibration purposes.

Following our numerical simulations, we have also performed an implementation on quantum processing units based on Trapped Ions~\cite{IonTrapQuantumComputing_review} and Superconducting Qubits~\cite{SuperconductingQubits_review}. More specifically, we use IonQ~\cite{Aria-1} and IQM~\cite{IQM_test} quantum processors. The experimental results show that our approach can be used as an effective tool to assess whether a given QPU satisfies the end user's needs, since it yields very precise information about what portion of the state space can be reached as the number of qubits grows. These experiments also raise the question about to which extent one can affirm that a given QPU is capable of simulating non-local states. 

The paper is organized as follows. First we give a brief overview of the Svetlichny and Mermin inequalities for multiple qubits and the notion of QRC. Next, we show the results of simulations of the violation of Mermin's and Svetlichny's inequalities for different numbers of qubits, shots, and noise levels, and compare the results of universal  vs non-universal sets of quantum gates. We also include the analysis of different entanglement measures and quantum magic. We then propose a certification protocol based on our theoretical developments, and perform some experiments on IonQ and IQM processors to show how our approach can be useful to assess the performance of different architectures. In the Discussion section, after making some remarks about the interpretation of the experimental results, we draw our conclusions and leave open some questions for future work.


\section{Detecting non-locality}

\begin{figure}[b!]
\centering
\includegraphics[width=\linewidth]{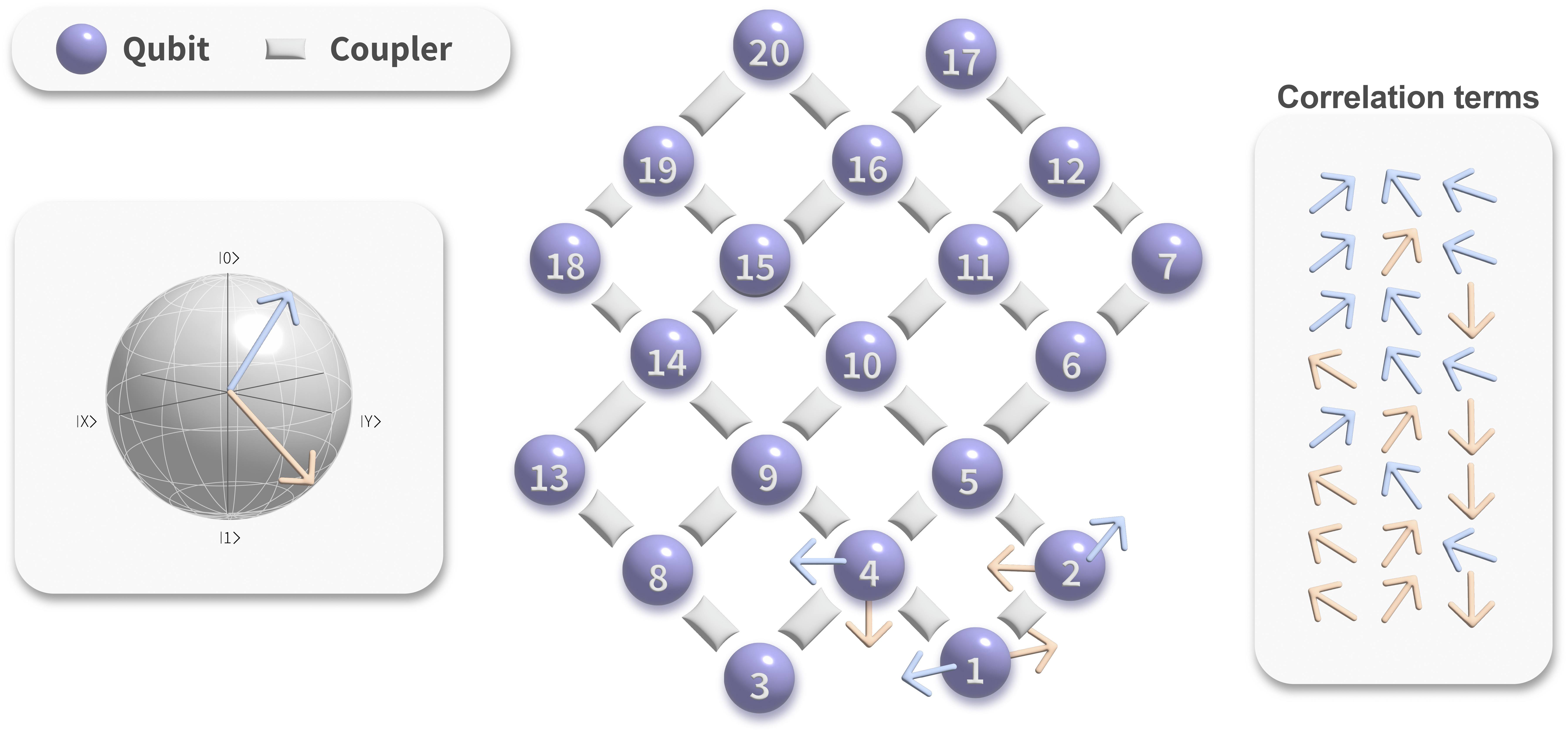}
\caption{\label{f:Experiment_IQM_Garnet_spin_orientations}Schematic representation of IQM Garnet's connectivity, where the three qubits Svetlichny inequality is to be tested on qubits $1$, $2$, and $4$. Each arrow represents a possible spin measurement on the chosen qubits. There are two directions per qubit, described by blue or yellow arrows. Thus, there are eight correlation terms in total (see Eqs.~\eqref{e:SvetlichnyA} and~\eqref{e:SvetlichnyB} in Appendix~\ref{s:MerminAndSvetlichnyExplained}), which are schematically depicted as eight orientation choices on the right box.}
\end{figure}

Since the celebrated work of John S. Bell~\cite{Bell_Aspect_2004}, the notion of quantum non-locality entered into the stage of quantum physics and gave place to heated foundational debates. This phenomenon, understood as a violation of local realism, is mathematically characterized by the lack of a probability distribution that simultaneously reproduces the observed correlations and meets a certain separability condition~\cite{Bell_Non-Locality}. The failure of local-realism is expressed in the violation of a family of inequalities involving correlations among the different parties of the system.

For multi-component quantum systems, different types of non-locality inequalities exist. This work focuses on the Mermin~\cite{Mermin_1990} and Svetlichny~\cite{Svetlichny_PRD,Seevinck_Svetlichny_PRL} inequalities (see Appendix~\ref{s:MerminAndSvetlichnyExplained} for details). A key difference is that violating a Svetlichny inequality implies the system exhibits genuine multipartite non-locality. This means that there exists no bipartition of the system for which the observed correlations can be explained by any local-realistic model. In this sense, a violation of the Svetlichny inequality allows to discard what is known as \textit{hybrid local realism}.

\begin{figure}[t]
\centering
\includegraphics[width=\linewidth]{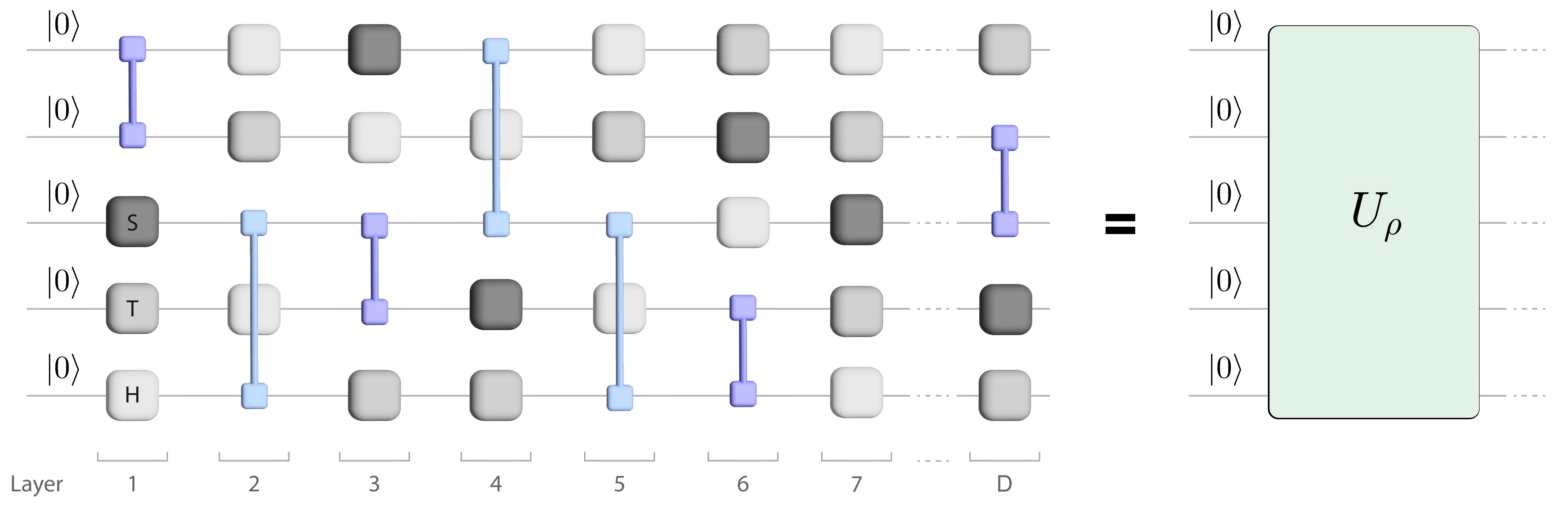}
\caption{\label{f:Circuit}Schematic representation of the random choice of gates on a five qubits circuit. We used the Amazon Braket SDK~\cite{braket} to develop a function that generates quantum random circuits (QRC). We first fix a set of elementary gates (as for example, those of the Clifford $+$ \texttt{T} set). Then, on each layer of the circuit, we make a random selection of gates and apply them to randomly chosen qubits. In the schematic diagram, we show different types of one-qubit gates (boxes indicating \texttt{H}, \texttt{T} and \texttt{S} gates), together with two-qubit entangling gates (such as, for example, \texttt{CNOT}). 
After the action of each random circuit, the target qubits are left in an output state $\rho$ that depends on the circuit. Each circuit is equivalent to a random unitary operator \(U_{\rho}\).} 
\end{figure}

Given a quantum system of multiple qubits prepared in the state $\rho$, we consider local spin measurements on each qubit. To fix ideas, in Fig.~\ref{f:Experiment_IQM_Garnet_spin_orientations} we show a scheme of one of the experiments we performed on an IQM (\textit{Garnet}) QPU. In order to study the violation of the three-qubits Svetlichny inequality, one must compute the correlations among spin measurements associated to a chosen set of qubits. The Svetlichny inequality consists in a linear combination of such correlations, each term containing local spin measurements in different orientations  (see Eqs.~\eqref{e:SvetlichnyA} and~\eqref{e:SvetlichnyB} in Appendix~\ref{s:MerminAndSvetlichnyExplained}). These terms are schematically indicated on the right of Fig.~\ref{f:Experiment_IQM_Garnet_spin_orientations}. 

By applying quantum random circuits (QRC) on a chosen set of qubits, we can prepare them in different random states. QRC are used in different areas of quantum information science~\cite{Bouland2019}. In order to give a general idea, consider a scheme of circuit generation as depicted in Fig.~\ref{f:Circuit}. At each stage of the protocol, random unitary gates are chosen and applied to the circuit. As a result, a random circuit is obtained. Remarkably, QRC are a promising tool to show quantum advantage~\cite{Arute_2019_Sycamore}. Here, we use them to show how quantum resources are distributed among the possible states that a quantum computer is able to generate. An ideal quantum computer should be able to reach all possible states of a system of $N$-qubits. A real quantum device is able to implement a given set of elementary quantum gates on a fixed set of qubits. By generating QRC with different depths, sets of elementary gates, and levels of noise, we study how robust the resource distribution is relative to the ideal case. Specifically, by comparing Clifford (\texttt{H}, \texttt{S} and \texttt{CNOT}) vs non Clifford (Clifford $+$ \texttt{T}) sets of gates, we highlight the differences in how quantum resources are distributed between a non-universal and classically simulable quantum computing model, vs a universal one.

After the application of a particular random circuit, one ends up with a particular quantum state $\rho$, which will be used to perform a nonlocality test. For that goal, one must measure correlations between different combinations of local spin measurements, and use them to compute the value of the Svetlichny inequality (see Eq.~\eqref{e:SvetlichnyA}). As explained in Appendix~\ref{s:MerminAndSvetlichnyExplained}, we chose the angles in such a way the quantity defined by Eq.~\eqref{e:SvetlichnyA} is maximal for the chosen target state. In what follows, we refer to this quantity as the \textit{violation level} of a given state ---even if no violation might be present. The local spin measurements are simulated using local spin rotations as indicated in Fig.~\ref{f:Diagrama_sve}. The implementation of this procedure in a real quantum processor will depart from the ideal case, due to the presence of noise and imperfections. In this way, one can quantify to which extent the processor can reproduce the correlations of a state associated to a quantum random circuit. The procedure for testing Mermin or Svetlichny inequalities for an arbitrary number of qubits is completely analogous to the example discussed in this section.

In order to assess the performance of a quantum device as a whole, in the rest of this work we use the fundamental physics test provided by the Mermin and Svetlichny inequalities combined with the implementation of QRC. Under ideal conditions, by using thousands of instances of quantum circuits randomly generated, one should obtain a fair sample of the quantum state space as a whole. For each circuit thus generated, we quantify the violation level of the associated state. This allows us to build histograms indicating how this magnitude is distributed throughout the quantum state space. For completeness, we have also performed a similar study for other quantities, such as entanglement measures and quantum magic.

\section{Numerical simulation of the violation of non-locality inequalities}

\begin{figure}[ht!]
\centering
\includegraphics[width=\linewidth]{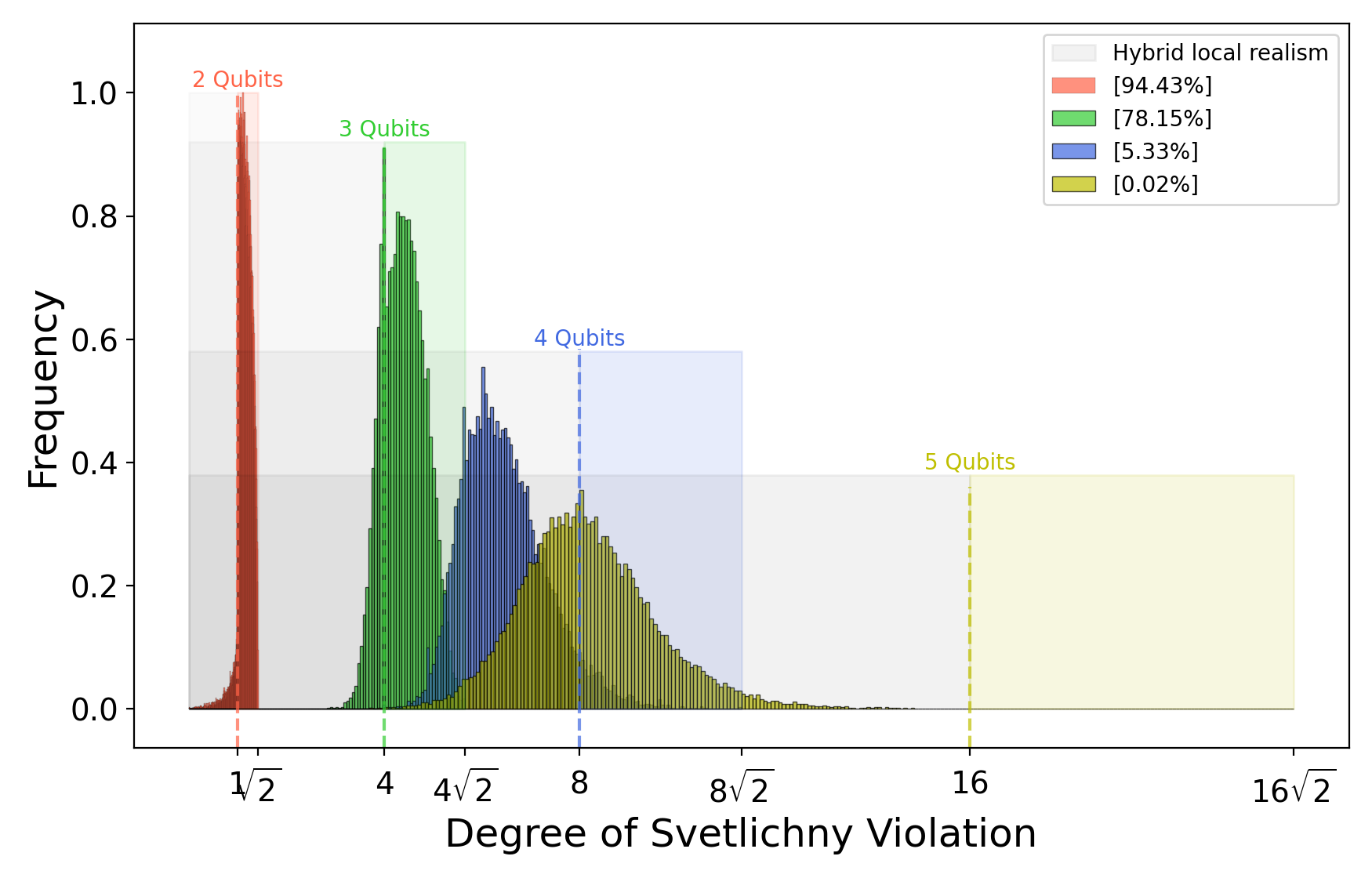}
\caption{\label{f:Histograms_condensed}Histograms of maximal violation of the Svetlichny inequality from three to five qubits, using $100{,}000$ randomly generated states. For the sake of completeness, we have also included the histogram corresponding to the violation of the CHSH inequality (two-qubit case). Different qubit numbers are indicated with colors. The vertical dashed lines denote the limit discarding hybrid local realism for each case. The percentage of states violating the inequality for each case is indicated in the top right. The universality of the Clifford $+$ \texttt{T} set implies that the above random circuits should approximate those generated with unitary matrices distributed according to the Haar measure\footnote{The Haar distribution is a translation-invariant probability measure. Intuitively speaking, it is a uniform probability distribution in the space of unitary matrices.}. We have obtained very similar results using both methods, which serves as a check of the correctness of our results (see Fig.~\ref{f:Histograms_Svetlichny} in Appendix~\ref{s:Histograms}).}
\end{figure}

We now present the numerical results. Fig.~\ref{f:Histograms_condensed} displays the histograms of the violation levels of the Svetlichny inequality for states generated using QRC from three to five qubits.

\begin{widetext}
\onecolumngrid
\begin{figure}[h]
    \centering
    \includegraphics[width=0.85\textwidth]{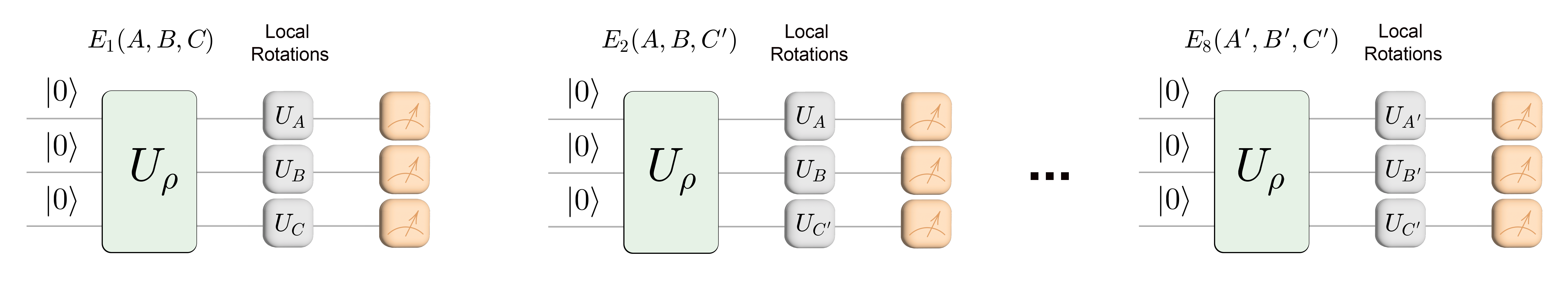}
    \caption{\label{f:Diagrama_sve}Three-qubits construction of the Svetlichny inequality. From the prepared state \(U_\rho\) (see Fig.~\ref{f:Circuit}), local rotations \(U_A,U_B,U_C\) and their primed settings define the eight tripartite correlators \(E_i\) used to build the Svetlichny operator \(O_3\) and evaluate \(S_3=\langle O_3\rangle\). Each local rotation of a qubit with the subsequent measurement in the computational basis simulates a local spin measurement.}
\end{figure}
\end{widetext}
\twocolumngrid


\begin{figure}[hb!]
\centering
\includegraphics[width=\linewidth]{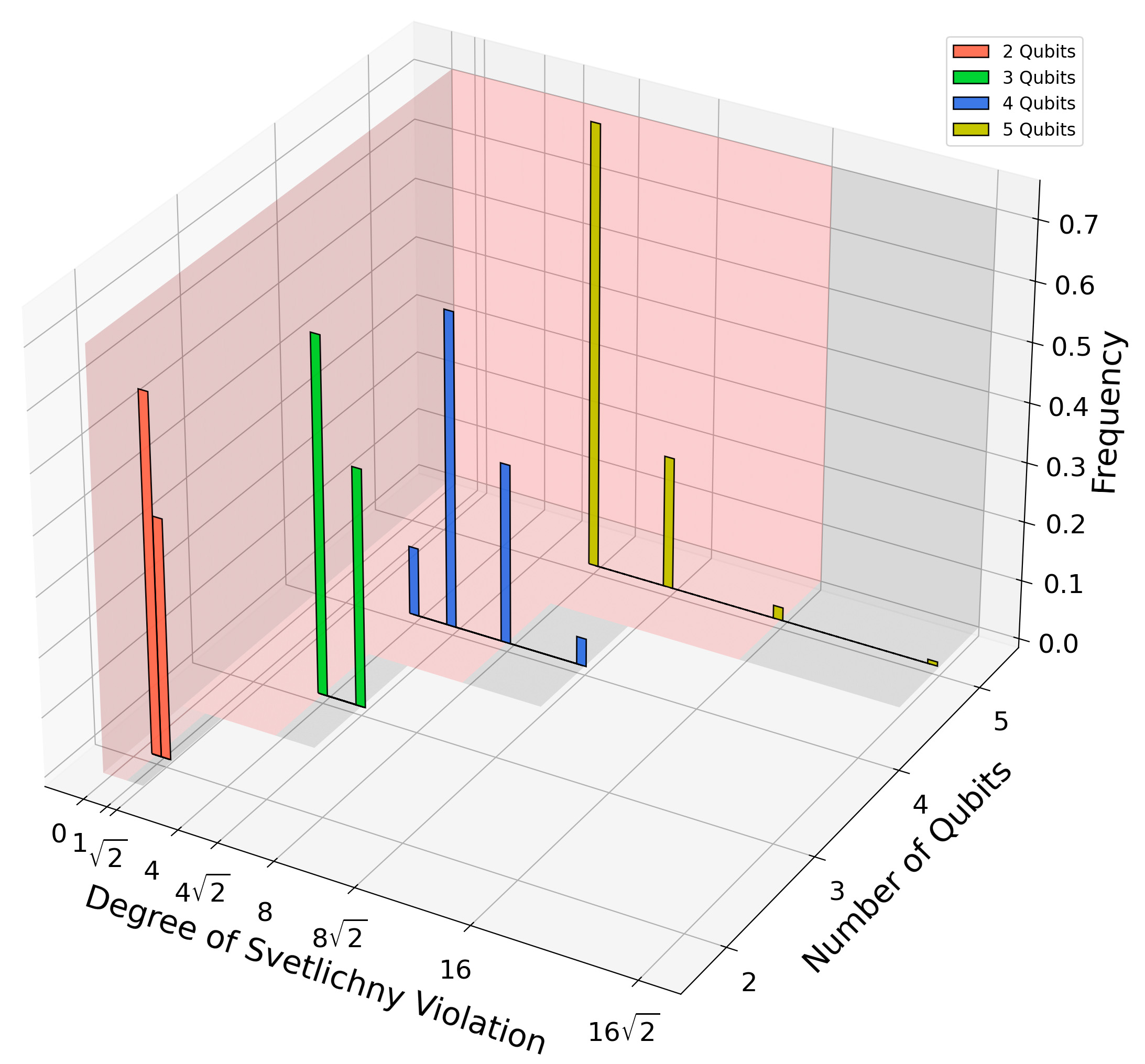}
\caption{\label{f:Clifford_Svetlichny}Histograms for the maximal violation degree of the Svetlichny inequality for states generated with random circuits using only Clifford gates from three to five qubits. For completeness, we have also included the histogram corresponding to the CHSH inequality. We display the relative weights of the values obtained with $100{,}000$ instances. Differently from the universal gate set case of Fig.~\ref{f:Histograms_condensed}, the distribution of degrees of nonlocality is now concentrated in certain specific values. Notice that, in all cases, there exist states reaching the maximal violation value allowed by quantum theory. The existence of such states holds for an arbitrary number of qubits, given that the generalized GHZ states can be produced using only Clifford resources.}
\end{figure}

\noindent In Fig.~\ref{f:Clifford_Svetlichny} we show the results obtained for quantum random circuits that use only Clifford gates. The comparison between the studied Clifford $+$ \texttt{T} and Clifford sets of gates indicates that one of the main differences comes from the way in which resources are distributed. In both cases, we obtain the maximum possible value of violation.
By Gottesman–Knill theorem~\cite{Cuffaro_Gottesman_Knill_Theorem}, this means that even a quantum device which is classically simulable can produce states with maximal non-locality. This holds for an arbitrary number of qubits, given that the GHZ states can be prepared using only Clifford resources. But, in a universal device, the different degrees of non-locality are homogeneously distributed among all possible values available in the full quantum state space. This is in line with previous works (compare with the results of~\cite{monchietti2023}). It is important to remark that the states generated using the Clifford set have a very different geometrical structure with regard to those generated with Clifford $+$ \texttt{T} (compare Figs.~\ref{f:Histograms_condensed} and~\ref{f:Clifford_Svetlichny}). While in the latter the states generated tend to cover the entire set of quantum states (and it is very unlikely to obtain a given state twice), in the former it is indeed likely to obtain repeated states (when the Clifford circuits are chosen at random), since the number of Clifford states is finite for any number of qubits and quite small until four qubits. Thus, for the Clifford set, the term ``random'' only applies to the way in which the gates of the elementary set are chosen to form each circuit (i.e., following the procedure indicated in Fig.~\ref{f:Circuit}), and should not be confused with that of a uniform distribution which applies when using random unitaries (and which the universal set resembles). All possible states generated using Clifford gates can be actually computed~\cite{Aaronson_Stabilyzer_Circuits_2004}, and we have done that up to five qubits. We have checked that up to four qubits, the statistical distribution of degrees of violation values associated to all possible Clifford states retain the same relative weights as those presented in the histograms shown here. For five qubits, there are too many Clifford states, and we have opted to show the relative weights obtained using random circuits. Furthermore, the states associated with the Clifford set exhibit a rich and intricate geometric structure, the detailed investigation of which will be the subject of a separate study.

\begin{figure*}[t]
\centering
\includegraphics[width=0.84\textwidth]{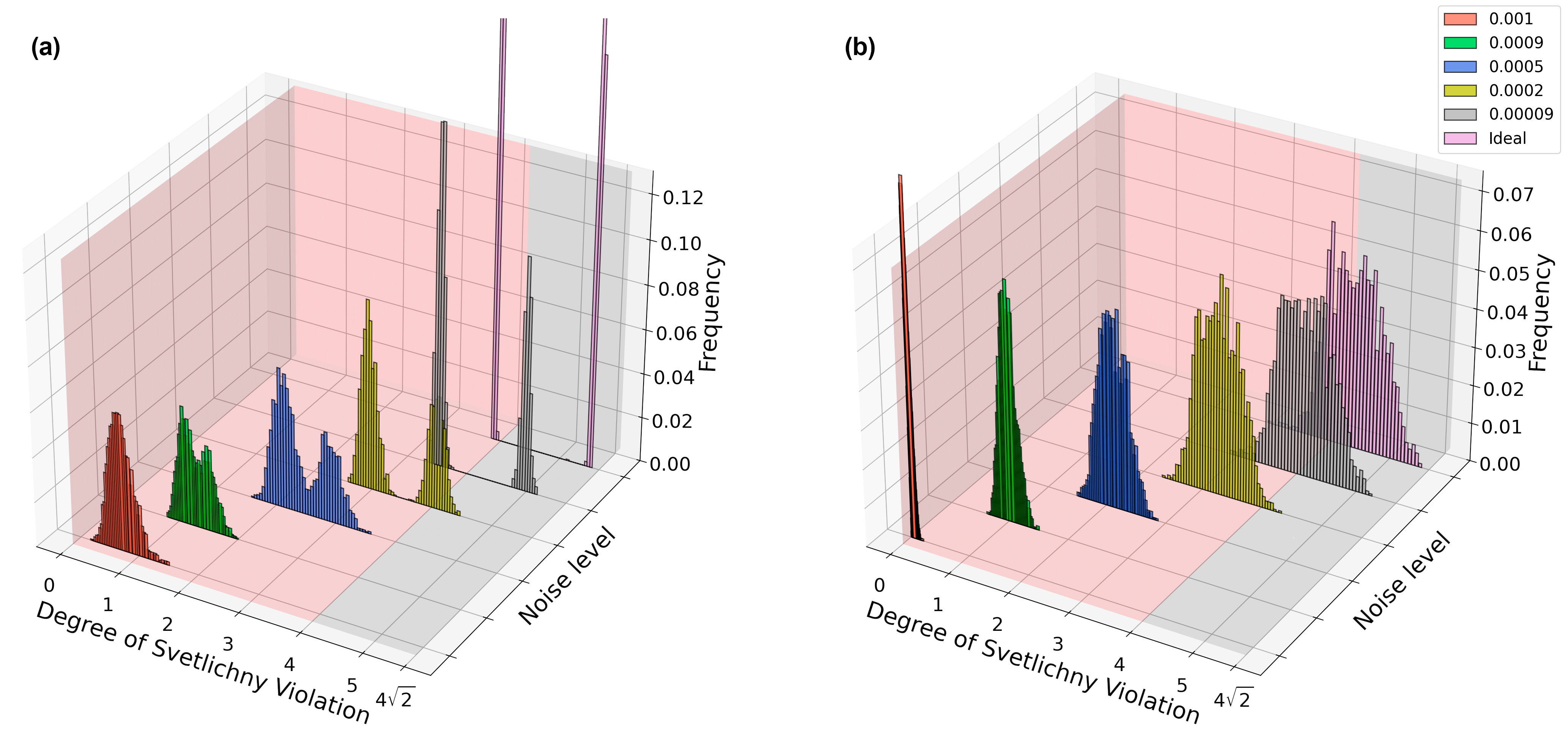}
\caption{\label{f:3qSvetlichny_noise}Degree of Svetlichny violation histograms for different levels of depolarizing noise using three-qubit circuits composed of gates taken from the (a) Clifford and (b) Clifford $+$ \texttt{T} sets. The different values of the noise parameter are shown on the top right corner. Comparison between the ideal histograms (in pink above) and the experimental ones can serve as the basis of a benchmarking protocol.}
\end{figure*}

In Fig.~\ref{f:3qSvetlichny_noise} we show the violation levels obtained for three qubits for different degrees of depolarizing noise. Using this simple noise model serves to illustrate our point: how the nonlocality histograms are deteriorated under the imperfections of quantum devices. By augmenting noise Fig.~\ref{f:3qSvetlichny_noise}, one sees that the ideal histograms ``move to the left" in both cases. By comparing how an empirical histogram departs from the ideal case, one could certify how robust the resource's distribution is in a concrete quantum device. The above analysis can be considered then as the basis of a benchmarking protocol that we will discuss below.

\begin{figure}[b!]
\centering
\includegraphics[width=\linewidth]{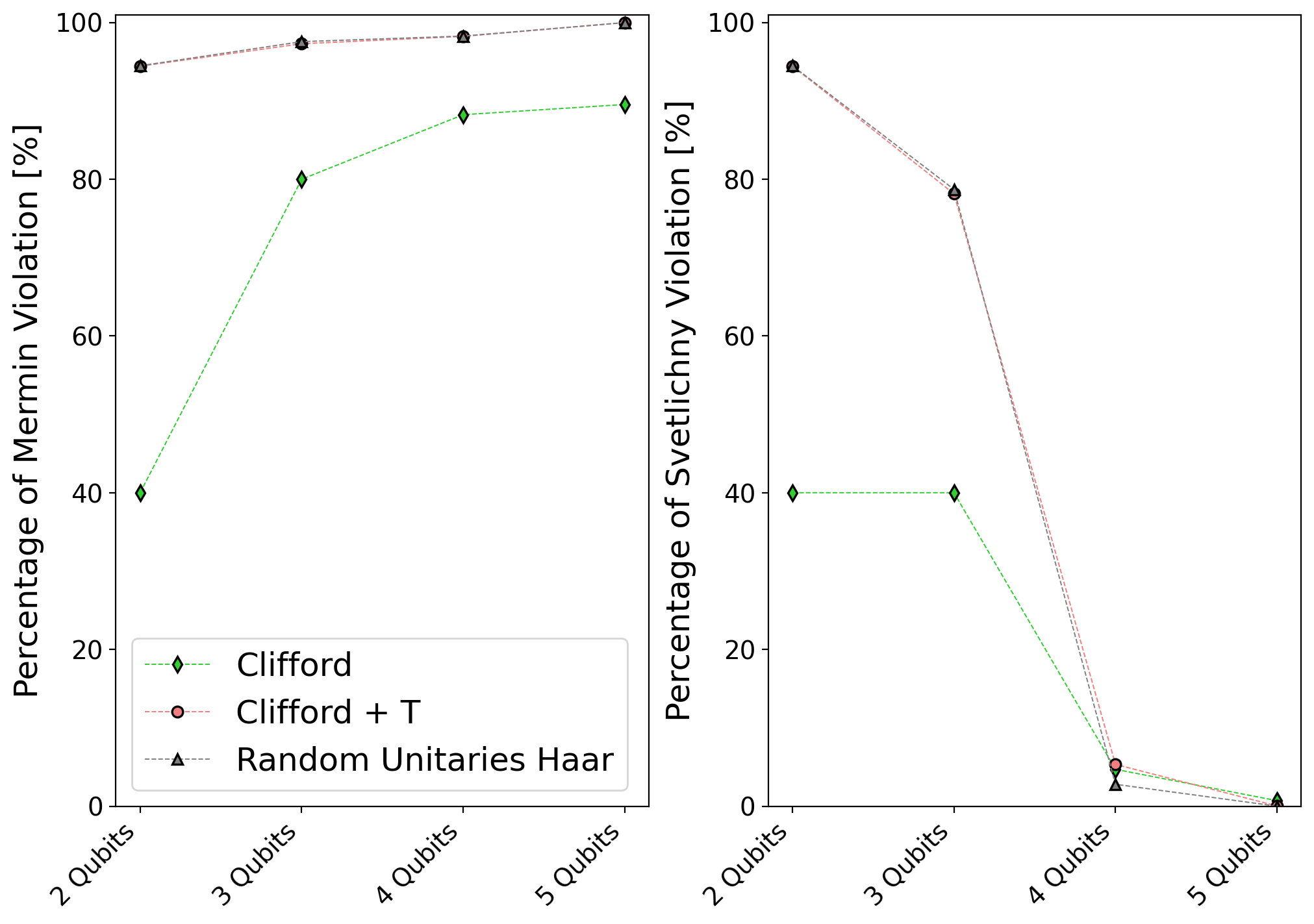}
\caption{\label{f:S_and_M_Violations}Violation fraction for Mermin and Svetlichny inequalities as a function of the number of qubits using Clifford gates, Clifford $+$ \texttt{T} gates, and random unitaries following the Haar distribution. For completeness, we also report the value corresponding to the CHSH inequality.}
\end{figure}

It is interesting to examine the violation fraction, defined as the percentage of generated states that violate a given inequality for some choice of angles (the angles of maximal violation depend on the state), as the number of qubits is increased (see Fig.~\ref{f:S_and_M_Violations} and Table~\ref{tab:Table_1}). For the Svetlichny inequality, we find that the violation fraction seems to tend to zero as the number of qubits grows. In other words, the probability that a randomly picked pure state displays genuine non-local multipartite correlations (in the Svetlichny scenario) becomes very small in that limit. On the contrary, for the Mermin inequality studied here, the violation percentage tends to remain very high as $N$ grows, indicating that the volume of the set of states displaying some form of non-local correlation is big (around $90\%$ of the whole state space). This is in agreement with the fact that the volume of the set of separable states tends to zero as $N$ goes to infinity and the \textit{typicality of nonlocality} phenomenon (see~\cite{Rosier2020}). It is important to recall here that entanglement and non-locality are different concepts \cite{Augusiak2015,Makuta2025}.

In particular, our results suggest that the volume of the set of states which do not display genuine multipartite nonlocality with regard to the observables appearing in the version of Svetlichny's inequality studied here, that is, Eq.~\eqref{e:SVL_N_Qubits}, does not seem to tend to zero as $N$ grows. But the fact we don't obtain a violation of that inequality for a given state, does not preclude the existence of other GMNL inequalities which might be violated by that same state. Therefore, based on our results, we cannot conclude that the volume of the set of GMNL states tends to zero. We will address this subtle question in future works.

\setcounter{table}{0}
\begin{table*}[t] 
  \centering
  \begin{tabular}{@{}c@{}|@{}c@{}|@{}ccc@{}|@{}ccc@{}|}
    \cline{2-8}
    & \textbf{$\ $CHSH$\ $} & \multicolumn{3}{@{}c@{}|}{\textbf{Mermin}}
    & \multicolumn{3}{@{}c@{}|}{\textbf{Svetlichny}} \\ \cline{2-8}
    & 2 & \multicolumn{1}{@{}c@{}|}{3} & \multicolumn{1}{@{}c@{}|}{4}
    & \multicolumn{1}{@{}c@{}|}{5} & \multicolumn{1}{@{}c@{}|}{3}
    & \multicolumn{1}{@{}c@{}|}{4} & 5 \\ \hline
    \multicolumn{1}{|@{}c@{}|}{\textbf{Clifford}}
    & \multicolumn{1}{@{}c@{}|}{\cellcolor[HTML]{EFEFEF}40\%}
    & \multicolumn{1}{@{}c@{}|}{\cellcolor[HTML]{EFEFEF}80\%}
    & \multicolumn{1}{@{}c@{}|}{\cellcolor[HTML]{EFEFEF}88.24\%}
    & \multicolumn{1}{@{}c@{}|}{\cellcolor[HTML]{EFEFEF}$\ $89.54\%$\ $}
    & \multicolumn{1}{@{}c@{}|}{\cellcolor[HTML]{EFEFEF}40\%}
    & \multicolumn{1}{@{}c@{}|}{\cellcolor[HTML]{EFEFEF}4.70\%}
    & \multicolumn{1}{@{}c@{}|}{\cellcolor[HTML]{EFEFEF}$\ $0.70\%$\ $} \\ \hline
    \multicolumn{1}{|@{}c@{}|}{\textbf{Clifford $+$ \texttt{T}}}
    & 94.43\%  & \multicolumn{1}{@{}c@{}|}{$\ $97.29\%$\ $}  & \multicolumn{1}{@{}c@{}|}{$\ $98.23\%$\ $} & 99.99\% & \multicolumn{1}{@{}c@{}|}{$\ $78.15\%$\ $} & \multicolumn{1}{@{}c@{}|}{$\ $5.33\%$\ $} & 0.02\% \\ \hline
    \multicolumn{1}{|@{}c@{}|}{\textbf{Haar}}
    & \cellcolor[HTML]{EFEFEF}94.48\%
    & \multicolumn{1}{@{}c@{}|}{\cellcolor[HTML]{EFEFEF}97.56\%}
    & \multicolumn{1}{@{}c@{}|}{\cellcolor[HTML]{EFEFEF}98.27\%}
    & \multicolumn{1}{@{}c@{}|}{\cellcolor[HTML]{EFEFEF}99.99\%}
    & \multicolumn{1}{@{}c@{}|}{\cellcolor[HTML]{EFEFEF}78.71\%}
    & \multicolumn{1}{@{}c@{}|}{\cellcolor[HTML]{EFEFEF}2.79\%}
    & \multicolumn{1}{@{}c@{}|}{\cellcolor[HTML]{EFEFEF}0\%} \\ \hline
    \rowcolor[HTML]{FFCCC9}
    \multicolumn{1}{|@{}c@{}|}{\textbf{LHV}}
    & 1 & \multicolumn{3}{@{}c@{}|}{1}
    & \multicolumn{3}{@{}c@{}|}{$2^{(N-1)}$} \\ \hline
    \rowcolor[HTML]{C0C0C0}
    \multicolumn{1}{|@{}c@{}|}{\textbf{$\ $Quantum bound$\ $}}
    & $\sqrt{2}$ & \multicolumn{3}{@{}c@{}|}{$2^{(N-1)/2}$}
    & \multicolumn{3}{@{}c@{}|}{$2^{(N-1)}\sqrt{2}$} \\ \hline
  \end{tabular}
  \caption{\label{tab:Table_1} The first three rows represent the violation fractions (percentage of states exceeding the classical local hidden variables (LHV) limit after angle optimization) for CHSH ($N=2$), Mermin and Svetlichny inequalities ($N=3\text{--}5$) across circuit ensembles produced in three different ways:
  using only Clifford gates, Clifford $+$ \texttt{T} gates, and random unitaries distributed according to the Haar measure. The last two rows report, for each case, the LHV bounds and the corresponding maximal violation value predicted by quantum theory (here abbreviated as \emph{quantum bound}). These were used
  as reference thresholds in the histograms. Notice that the physical meaning of the LHV-limit is different in Mermin and Svetlichny inequalities, since in the former, it refers to plain local realism, while in the latter, it refers hybrid local realism. Percentages are obtained from noiseless simulations with $100{,}000$ random-circuit instances per setting. Notice that for the CHSH inequality case, we have used a different normalization from the standard one, for which LHV $= 2$ and quantum bound $=2\sqrt{2}$ (which is known as \emph{Tsirelson's bound}~\cite{Tsirelson_1987} in the literature).}
\end{table*}


Up to now, we have focused on non-locality, quantified as the violation levels of Mermin and Svetlichny inequalities. For the sake of comparison, it is also interesting to analyze what happens with entanglement and quantum magic (see~\cite{macedo2025witnessingmagicbellinequalities} for a recent discussion on the connection between the latter and nonlocality). 

\begin{figure}[ht!]
\centering
\includegraphics[width=\linewidth]{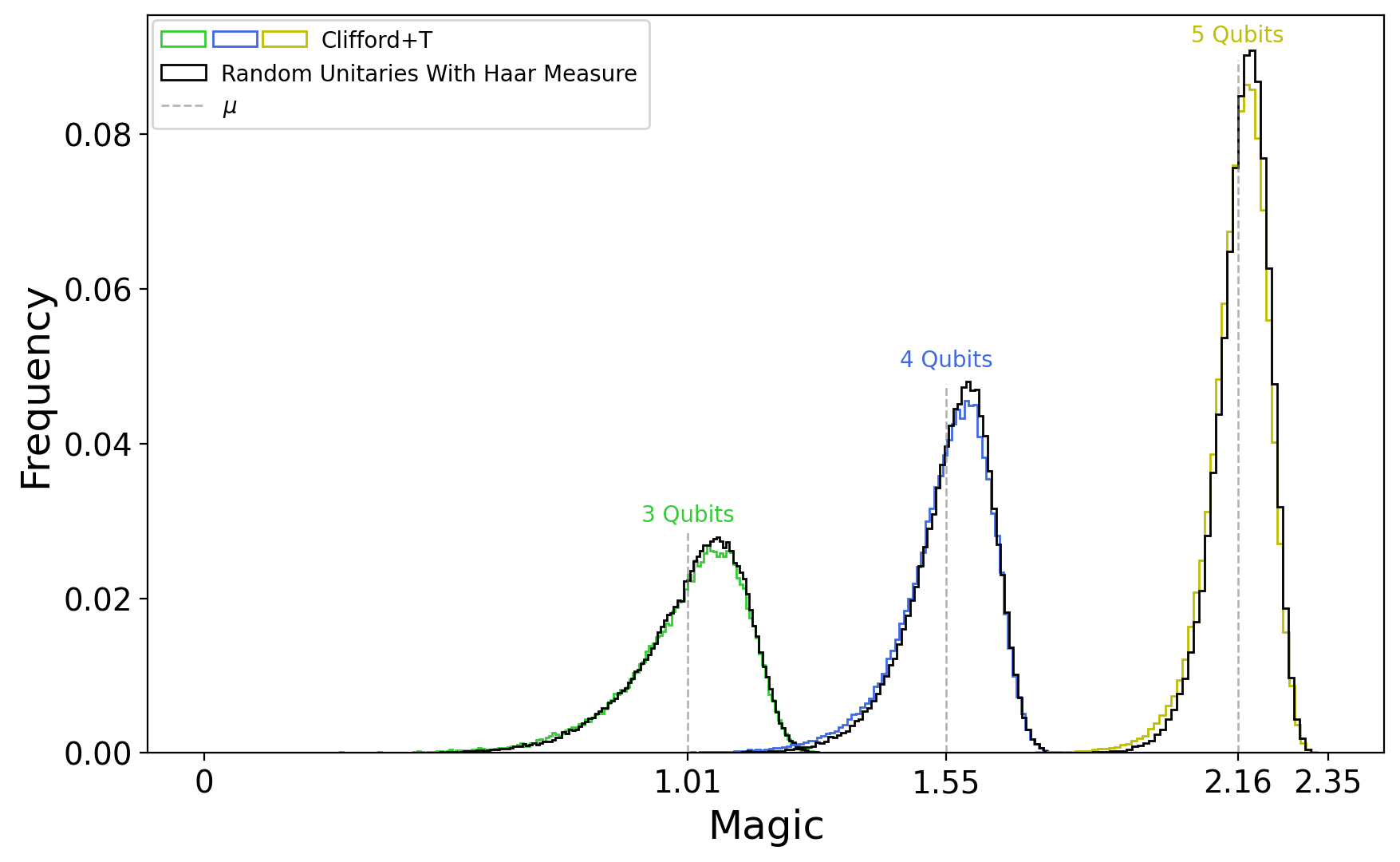}
\caption{\label{f:Quantum_Magic}Distribution of Quantum Magic across $100{,}000$ randomly generated circuits  for three to five-qubit systems. The circuits were constructed utilizing gates from the Clifford $+$ \texttt{T} set and random unitaries. As expected, for circuits constituted exclusively with the Clifford group, the quantum magic value is quantified as zero, hence its omission from the display. It is important to stress here the relevance of the depth of the circuits generated using the Clifford $+$ \texttt{T} set. Our codes allow to set the depth of the generated circuits as a random parameter. We have done in the above plots. The latter option yields results which allow to reproduce those with random unitaries with more precision.}
\end{figure}

In Fig.~\ref{f:Quantum_Magic}, we show the behavior of quantum magic based on the stabilizer $\alpha$-Rényi entropy. Roughly speaking, this quantity describes how many non-Clifford gates are necessary to generate a given quantum state~\cite{Oliviero2022}. As the number of qubits is increased, the histograms seem to become narrower. 

\begin{figure*}[t]
\centering
\includegraphics[width=0.8\textwidth]{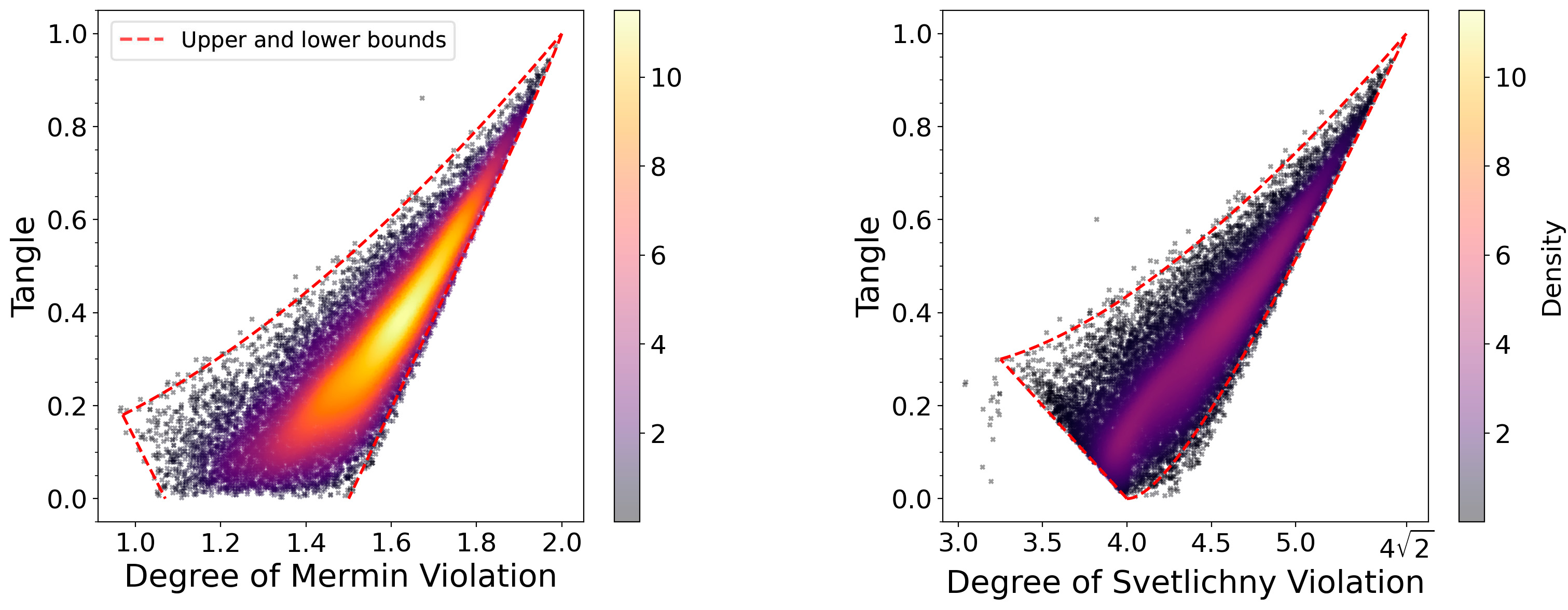}
\caption{\label{f:Mermin_vs_Svetlichny}Maximal violations of the Mermin (left) and Svetlichny (right) inequalities as a function of the three-tangle $\tau_{3}$ for $100{,}000$ three-qubit circuits using noiseless QRC with a universal set of gates. The red dashed curves indicate the upper/lower bounds associated with the family of states studied in Ref.~\cite{Emary_2004} (shown here for comparison).}
\end{figure*}

In Fig.~\ref{f:Mermin_vs_Svetlichny}, we plot the violation levels of Mermin and Svetlichny inequalities vs the tangle (see Appendix \ref{s:Appendix_entanglement_measures}) of $50,000$ three-qubit states generated using random unitary matrices. These plots extend the results presented in previous works (see for example~\cite{Emary_2004}).

In the remainder of this section, we refer to the figures contained in Appendix~\ref{s:Histograms}. A short overview of the entanglement measures used in this work can be found in Appendix \ref{s:Appendix_entanglement_measures}.

In Fig.~\ref{f:Distribution}, we show how different entanglement measures are distributed for the sets of random circuits studied in this work.
A comparison between histograms for Mermin and Svetlichny inequalities is displayed in Fig.~\ref{f:Histograms_Svetlichny} (also including the CHSH case). In Fig.~\ref{f:noise_behavior_all}, we show how the histograms deteriorate as noise is increased for the two-qubit case (CHSH inequality) and for the Mermin and Svetlichny scenarios in three-qubit systems. It is also important to remark that for the Clifford $+$ \texttt{T} set, the shape of the histograms depends on the depth of the circuits produced (i.e., a higher depth implies a better accuracy in the task of simulating arbitrary random circuits distributed according to the Haar measure). In Fig.~\ref{f:depth_variation_ct}, we show the violation levels obtained for Mermin and Svetlichny inequalities from two to five qubit systems varying the circuit's depths (without noise). For a higher depth, we find curves that resemble those of random unitaries. However, for lower depth levels, there are more chances of obtaining states with higher maximal violation levels (when the gates are randomly picked). This dependence could be of practical use when analyzing concrete devices. 

Fig.~\ref{f:noise_and_depth_behavior_all} gives a concrete snapshot of how the system behaves with regard to different levels of noise and depth. As expected, as noise grows, the violation fraction becomes smaller ---until eventually, it becomes zero. Notice also that, for a given noise level, there is a concrete depth for which the violation fraction becomes maximal. This is due to the fact that noise accumulates each time a quantum gate is applied. A low depth will give place to a poor family of states but, if the depth is too high, noise starts to play a significant role, destroying all resources. 

It is also interesting to look at the violation levels for a higher number of qubits and the Clifford set only. The results are displayed in Fig.~\ref{f:all_qubit_numbers_Sve}. Using the Clifford set, it is easier to achieve maximal violation values using a few iterations of the random circuits algorithm. Again, we recover a pattern where the violation levels obtained for the different states are concentrated in specific values. For a better visualization of this fact, we have plotted the list of values obtained for different qubits numbers in Fig.~\ref{f:all_qubit_numbers_Sve}. The violation level concentration is the most salient difference between universal and non-universal sets of gates studied in this work. Remarkably, the number of peak values obtained seems to be proportional to the number of qubits.

\section{\label{s:Certification}Benchmarking quantum processors}

The results obtained in the previous sections indicate that there is a sharp and quantifiable difference between, first, universal vs non-universal sets of gates, and second, ideal (noiseless) and non-ideal quantum processors. By comparing the ideal vs experimentally obtained histograms of violation degrees, one can have a global measure of to what degree a quantum processor is able to generate the whole quantum state space. This suggests a methodology for testing the performance of a device, following the steps below: 

\begin{enumerate}
    \item \textbf{Input}: Specify the number $N$ of quantum random circuits generated, number of qubits to be used $Q$, depth of the circuit $D$ and number of shots $S$ to be used.
    \item \textbf{Implementation on the Device} Using the native gates of the device, generate $N$ quantum random circuits ($S$ shots per circuit) and perform the measurements needed for the chosen resource to be tested. For the case of a multipartite Bell-type inequality, we should perform the measurements related to the operator associated to that observable.
    \item \textbf{Comparison with ideal device}: Compare the obtained results with the ideal ones (using the same set of gates). It could be to perform a histogram, compute the probabilities associated, or use some entropic measure to compare. 
\end{enumerate}


\section{\label{s:Implementation_In_QPUs}Implementation on quantum processors}

Here we show an example of implementation in IonQ (\textit{Aria-1}) and IQM (\textit{Garnet}) processors. Since implementing thousands of states in a real quantum processor is resource demanding, here we focus on a variant of the steps $1-3$ mentioned in the previous subsection. During the third step, instead of implementing all the randomly generated circuits, we proceed as follows. Focusing on the three-qubit case, we produce several instances of random circuits in classical hardware to produce a violation degree histogram (as in the previous section), and divide the horizontal axis interval $I = [4, 5.6]$ (the violation values range) in subintervals $I_j$ of length $0.2$ (giving place to ten groups of values). Next, from each interval $I_j$, we look at the set $\Delta_{I_{j}}$ formed by the circuits whose corresponding states have a maximal violation value in the interval $I_j$. From the set $\Delta_{I}$, we pick a representative circuit using the criteria of minimizing the number of gates and the number of entangling gates it uses (\texttt{CZ} in the IQM (\textit{Garnet}) and \texttt{MS} in the IonQ (\textit{Aria-1}) case). Following this procedure, we obtain a set of ten circuits that are representative of the violation values associated to the chosen subintervals of the horizontal axis of the histogram.

\begin{figure}[t]
\includegraphics[width=\linewidth]{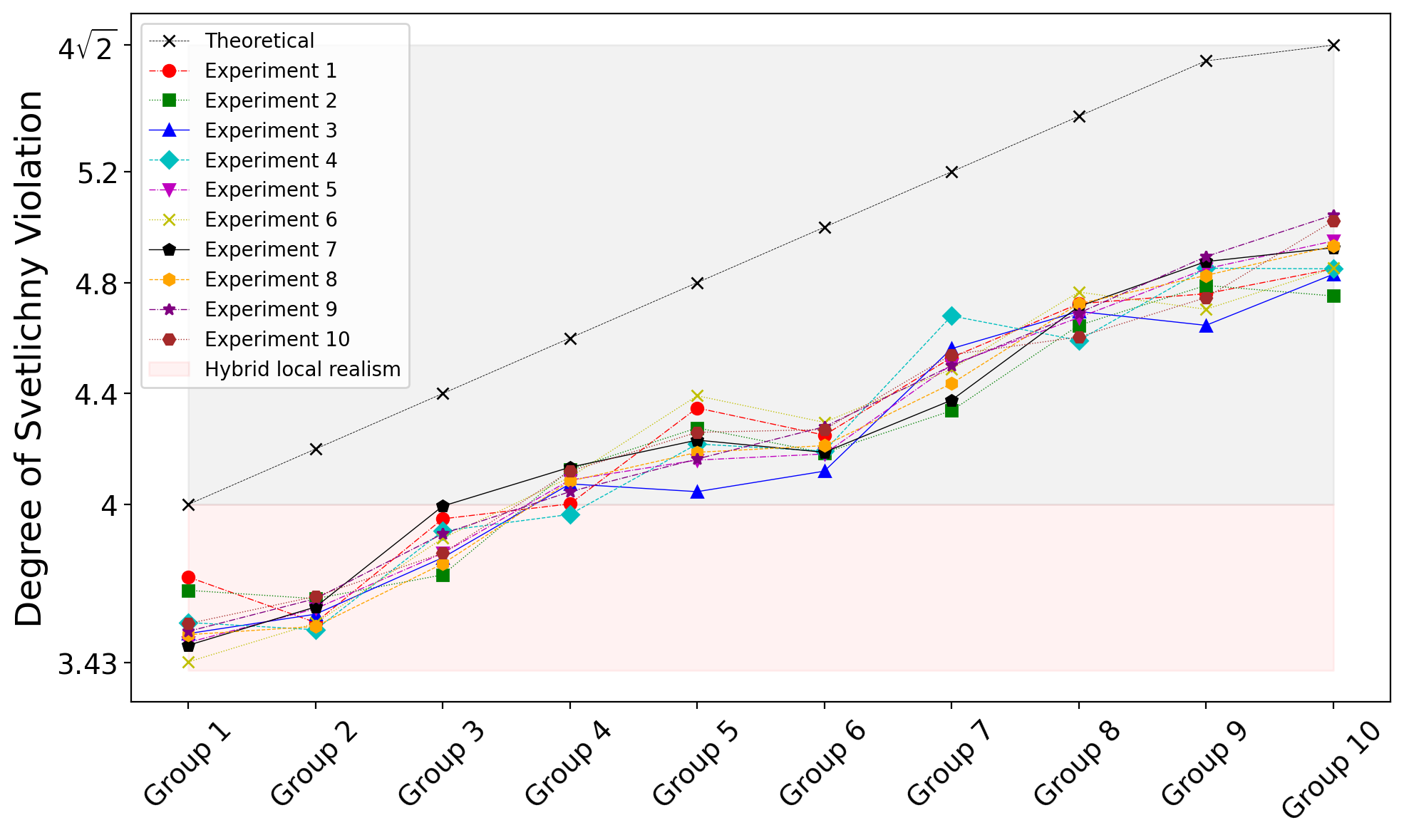}
\caption{\label{f:Experiment_IQM_Garnet_stats}Using IQM (\textit{Garnet}), we have simulated the violation degree of the Svetlichny inequality for a set of ten suitably chosen random circuits, which represent different parts of an ideal histogram. Each violation experiment was repeated ten times.}
\end{figure}

\begin{figure}[h!]   
\includegraphics[width=\linewidth]{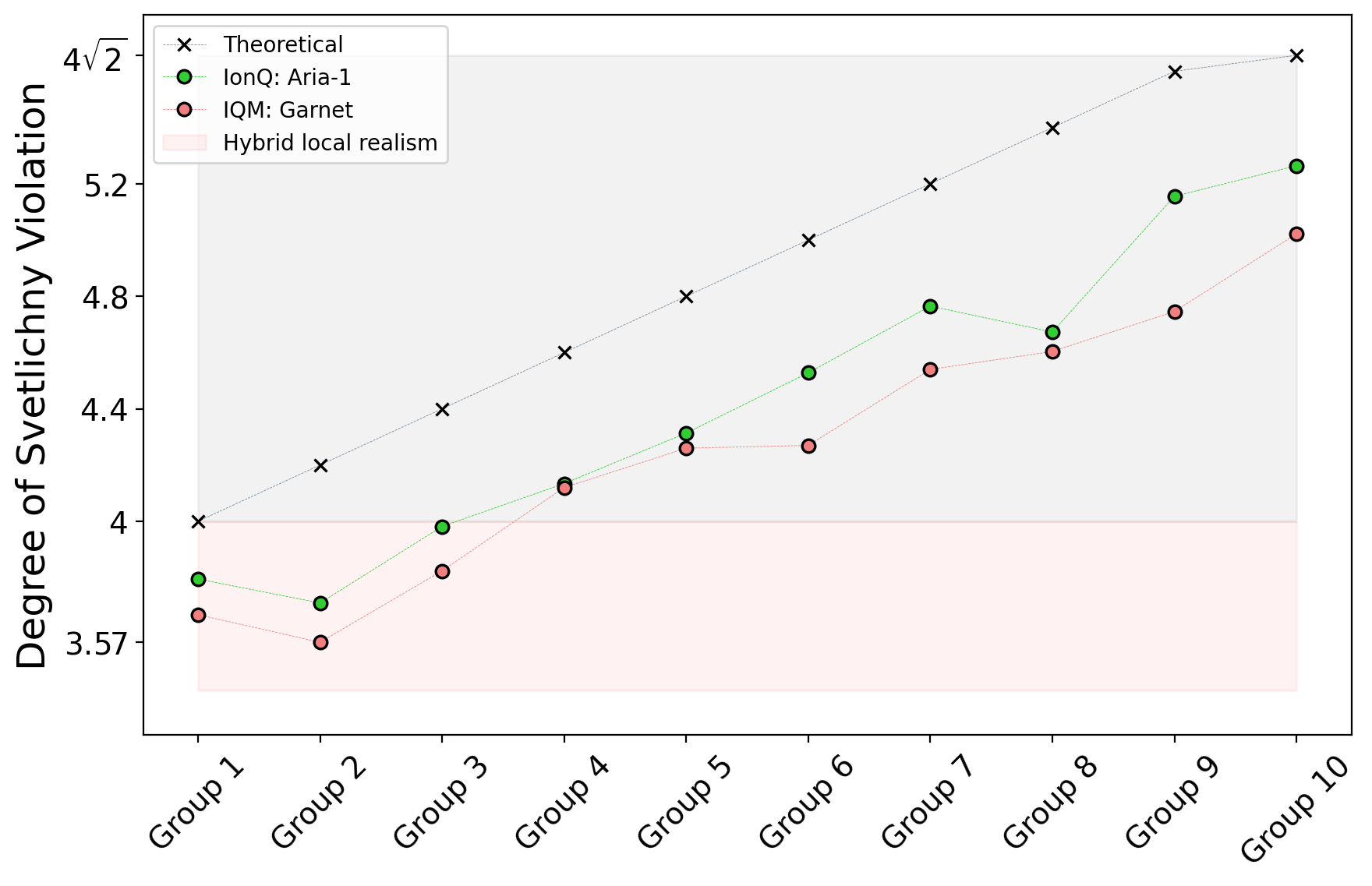}
\caption{\label{f:IonQ_vs_Garnet}Comparison between IonQ (\textit{Aria-1}) (green dots) and IQM (\textit{Garnet}) (red dots) in the task of simulating the violation of the Svetlichny inequality. The ideal violation values are indicated with black crosses. For each interval, we chose a representative state which uses a minimal number of \texttt{MS} and \texttt{CZ} gates respectively (and a minimum global number of gates). This move allows choosing circuits which are optimal for each QPU. For each circuit representing a quantum state we compute the values of angles of orientation of spin measurements with maximal violation of the inequality, and use them to simulate it on each QPU. Comparison between theory and experiment, and between different QPUs, can be used as a performance characterization.}
\end{figure} downloading the PDF and viewing it in a dedicated PDF reader, check for and correct any formatting errors accor

For producing the set of random circuits in classical hardware, we modified the function used in the previous section to produce circuits of variable length taking into account the connectivity of each processor, and using only native gates (\texttt{PRX} and \texttt{CZ} for IQM (\textit{Garnet}), and \texttt{GPi}, \texttt{GPi2} and \texttt{MS} for IonQ (\textit{Aria-1})). Local spin observables are implemented as proper rotations with \texttt{PRX} or \texttt{GPi}-\texttt{GPi2} gates depending on the processor (the theoretical spin orientation angles obtained for the synthetic data can be easily translated in terms of circuits formed by those gates). A schematic picture of the IQM (\textit{Garnet}) connectivity and the different spin directions used in a test of a Svetlichny violation is shown in Fig.~\ref{f:Experiment_IQM_Garnet_spin_orientations}. The results of simulating the Svetlichny inequality violation in the QPUs for the chosen representative circuits is displayed in Fig.~\ref{f:IonQ_vs_Garnet}, together with their corresponding theoretical values of maximal violations. Comparison between theory and experiment gives an idea of how much the histogram of violation values differs from the ideal case of a noiseless device. It also allows for a performance comparison between different QPU architectures, given that the \textit{shape} of the noiseless violation histograms is independent of the chosen universal set of gates.

It is also instructive to see what happens when the same violation experiment is repeated several times. In Fig.~\ref{f:Experiment_IQM_Garnet_stats} we show a plot with ten repetitions of the simulation of the Svetlichny inequality in IQM (\textit{Garnet}) for each state of Fig.~\ref{f:IonQ_vs_Garnet}. The results obtained indicate that while a violation of the Svetlichny inequality is clearly obtained, the results are far from the ideal case (compare these results with those of Fig.~\ref{f:Histograms_Svetlichny}).

\begin{figure}[h]
\includegraphics[width=\linewidth]{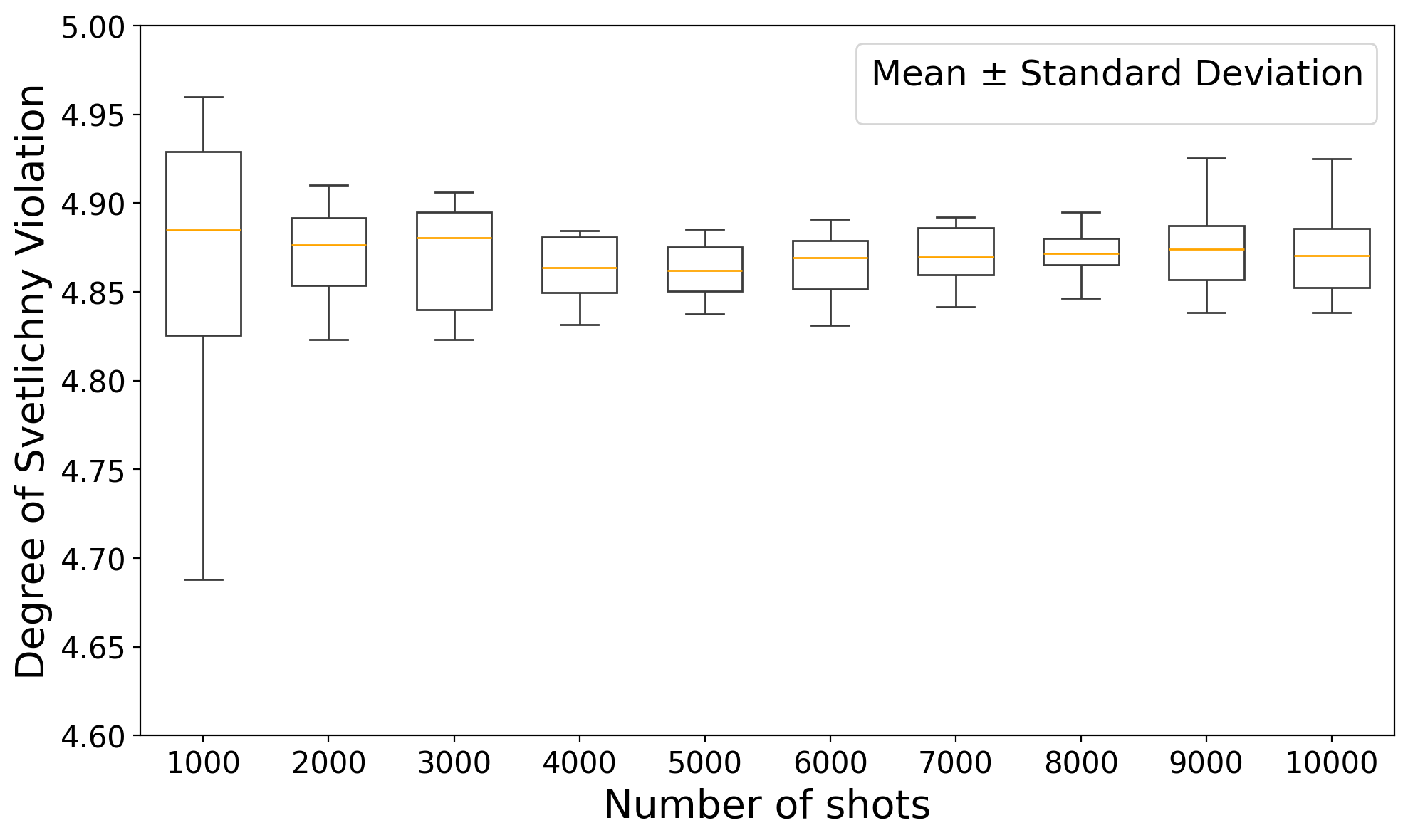}
\caption{\label{f:Garnet_3q_Svetlichny_shots}Plot of the violation values obtained for the violation of the Svetlichny inequality for a randomly chosen circuit associated to a state of maximal violation in IQM (\textit{Garnet}) as a function of the number of shots. We have prepared ten repetitions of each experiment.}
\end{figure}

As the results of each instance of a Svetlichny violation fluctuate, we have also performed an analysis of the dependence on the number of shots in IQM (\textit{Garnet}). The results are displayed in Fig.~\ref{f:Garnet_3q_Svetlichny_shots}. Clearly, the fluctuation tends to decrease as the number of shots grows. We find that using $1{,}000$ shots gives a good idea of the degree of violation attained by the QPU in the mean.

\begin{figure*}[]
\centering
\includegraphics[width=0.63\linewidth]{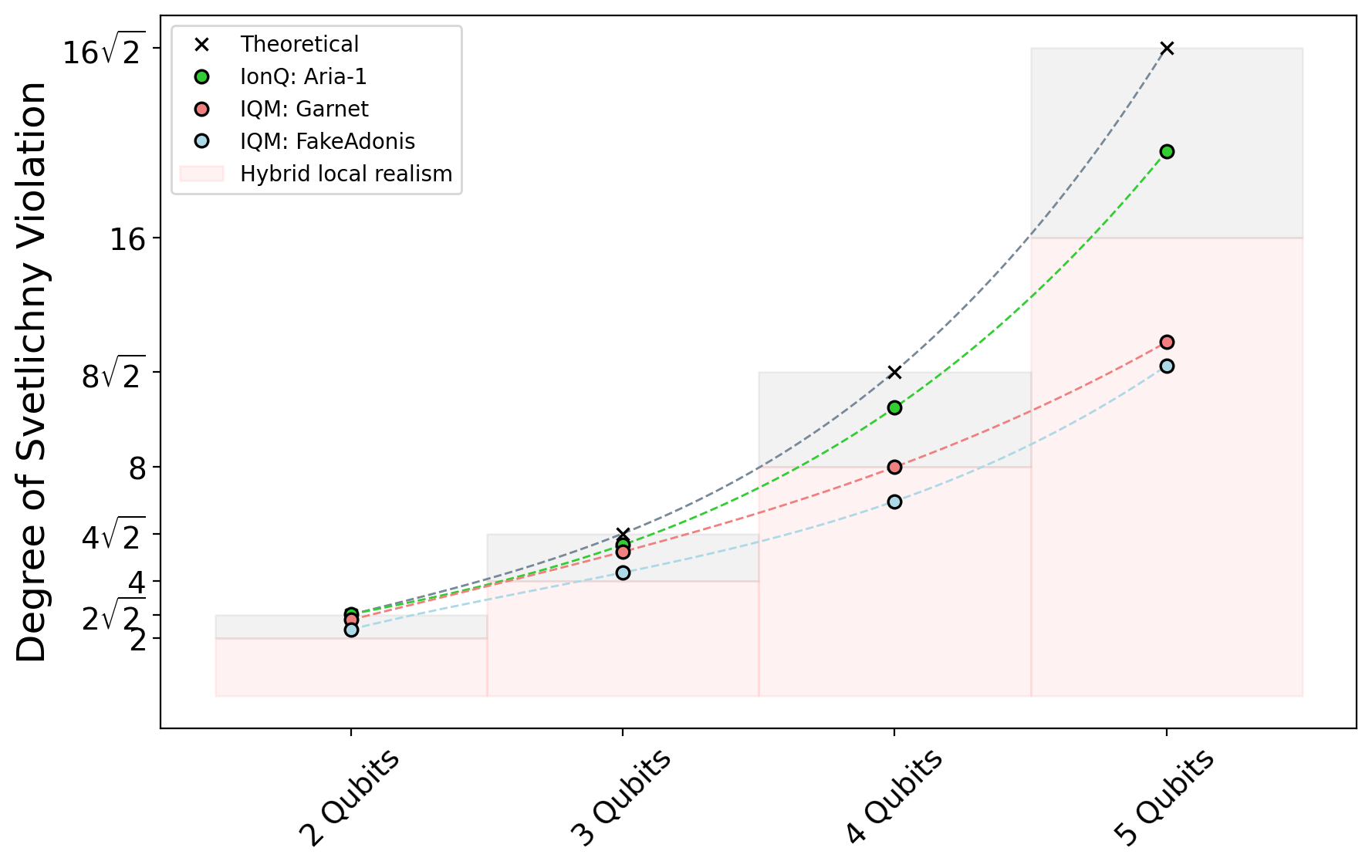}
\caption{\label{f:IonQ_Garnet_2-5_qubits}Degree of violation for GHZ states in IonQ (\textit{Aria-1})   and IQM (\textit{Garnet}) for two to five qubits. The statistics for the correlations of each experiment are computed with $1{,}000$ shots. For completeness, we additionally show the value corresponding to the CHSH inequality.}
\end{figure*}

Using IonQ's, IQM's optimization, and IQM's FakeAdonis classical simulator, we performed a final experiment with GHZ states for different numbers of qubits. The results are presented in Fig.~\ref{f:IonQ_Garnet_2-5_qubits}. We have obtained a simulation of the violation of the Svetlichny inequality up to five qubits with IonQ (\textit{Aria-1}).


\section{\label{s:Discussion}DISCUSSION}
\subsection{\label{s:Implementation_transpiled}Representation in terms of native gates}

It is important to make a digression about the interpretation of the simulation experiments presented in the previous section. Reproducing a non-locality test such as~\eqref{e:SVL_N_Qubits} in a quantum processor involves first the preparation of the state to be tested, and second, the implementation of the local observables out of which one computes the correlations appearing at the left-hand side of the inequality. Both tasks are carried out by finding suitably chosen quantum circuits. But the choice of these circuits is by no means trivial, as we explain in what follows.

When trying to simulate a state represented by the density operator $\rho$ in a quantum computer, one needs to implement a series of instructions $I = \left\{i_{1},i_{2},\cdots ,i_{n} \right\}$. Each instruction consists in the application of a quantum gate on the respective qubits. Let us call the state of the QPU resulting from those instructions $\rho^{ideal}_{I}$. If there were no imperfections present during the process, then, we should have $\rho=\rho^{ideal}_{I}$ (i.e., they should be the same operator). But in an actual experiment, due to imperfections at the level of gates implementation, there might be significant differences between the target state and the actual state obtained in the QPU. Besides, there is the problem of representation of I in terms of native gates. 

Therefore, the implementation of the instructions is automated in such a way as to optimize the task of creating a state as similar as possible to $\rho$. For that reason, the actual instructions implemented in the QPU will be denoted by a list $I' = \left\{i_{1}',i_{2}',\cdots ,i_{n}' \right\}$. While $I'$ might be quite different from $I$, if the optimization process is successful, the resulting state $\rho^{real}_{I'}$ will be very similar to $\rho$. Or equivalently, we end up with $\rho^{real}_{I'}\approx\rho^{ideal}_{I}$. Notice that the notation $\rho^{real}_{I'}$ refers to the actual state generated in the QPU (which might be in turn different from the ideal state $\rho^{ideal}_{I'}$ produced with noiseless instructions $I'$). The situation is similar when we implement measurements. Suppose that in order to emulate a certain set of spin measurements in a QPU we need to implement the list of instructions $M=\{m_1,m_2,...,m_l\}$. Let us call the resulting ideal state $\rho^{ideal}_{I+M}$ (i.e., the state that results from implementing noiseless instructions). Again, in an actual experiment, we would obtain $\rho^{real}_{I'+M'}$. Ideally, to perform the desired measurement in the state $\rho$, we should run the circuit with the instructions $I\cup M$ and sample on the computational basis. As performing a measurement in a basis different than the computational one is equivalent to performing a suitably chosen rotation and measuring in the computational basis if the optimization process works reasonably well, one concludes that $\rho^{real}_{I'+M'}$ is the best representative of the ideal process of preparing $\rho$ and measuring $M$. Notice also that in two different runs of the instructions $I\cup M$, one might obtain different circuits and, in principle, the chosen qubits might be even different. One could say that this is not relevant, because the goal is to reproduce the different copies of $\rho^{real}_{I+M}$ with the highest possible fidelity. It is important to notice that the above discussion can be applied mutatis mutandis to \textit{every} quantum experiment since every quantum experiment actually performed can be reduced to a list of operations on a physical system. The quantum systems used might be different (for example, different instances of photons or atoms), and the implementation of the procedures and measurements might slightly differ from one repetition to another. But the main goal of every experiment is to generate several copies of the ideal state and measurements (which are as similar as possible to it), in order to perform the required statistics. 

The results presented in the previous section should be interpreted taking into account the above considerations. One could say that what one does on each processor is to \textit{simulate} the Svetlichny inequality as exactly as possible in terms of quantum circuits. This observation reinforces the idea that these types of tests could be very useful to evaluate the overall performance of different sets of qubits of quantum processing units.

\subsection{\label{s:Contextuality}Genuine multipartite non-locality vs contextuality}

The experimental test of Bell inequalities was challenging due to the presence of possible loopholes: defects in the experimental implementation can be used to give explanations of the observed data in which the local-realism hypothesis can be maintained. After several years of huge efforts, the so-called loophole-free Bell tests were finally announced~\cite{Hensen_2015, Significant-Loophole-Free-2015, Strong-Loophole-Free-2015}. Taking into account how hard it was to reach that experimental degree of certification, one may wonder up to which extent it is realistic to affirm that genuine multipartite non-locality was produced in a given quantum processor. Indeed, in NISQ devices, qubits are usually very close, and the measurement of the different parties seems to be far from the conditions required for a loophole-free non-locality test. A quick thought indicates that it is unlikely that anyone can construct in the near future a quantum processor in which genuine multipartite non-locality can be tested in a loophole-free way. Therefore, we should be conservative about the certification of non-locality in current quantum processors (see also the discussion in~\cite{Cabello-Non-locality-benchmarking}, page 2). Still, the violation of Svetlichny inequalities in quantum processors can be considered a form of non-classicality, given that it discards a certain form of hidden variable models. Perhaps, it is safer to say that they represent a particular form of \textit{contextuality} in the following sense. For a certain family of observables (the local spin orientations with suitably chosen angles), there exists no global probability distribution satisfying certain constraints. Alternatively, one possible strategy is to work under the no-cross-talk assumption~\cite{Cabello-Non-locality-benchmarking}. In any case, it is reasonable to consider the violations presented in the previous section as a form of non-classicality, given that they imply that the obtained state presents genuine multipartite entanglement.\\ 
The above observations give place to a very interesting debate about the foundations of quantum physics that has technological implications. A proponent of the thesis that loophole-free certified non-locality is a necessary and sufficient condition for quantum advantage, will have to face the problem that it is extremely unlikely that a quantum processor will be ever able to take it for granted. This has implications on the development strategy of QPUs since, in the quest for quantum advantage, it would be necessary to move forward under the assumption that non-locality is actually present. Otherwise, efforts should be devoted to testing this feature in a loophole-free way, but this doesn't seem realistic with the extant technologies and experimental techniques. On the contrary, a defender of the thesis that quantum advantage is all about contextuality will still have a point in moving forward. One could always consider a quantum processor as a \textit{simulator} of genuine multipartite non-locality that uses contextuality as a resource. Is this enough for quantum advantage? Only time can tell the answer to this intriguing question. We will address this fascinating subject in more detail in a future work. With this short section, we just want to draw the reader's attention to the relevance of this problem, since it connects both, the foundations of quantum physics debates, and technological developments. We think that the results presented in this work contribute to a better understanding of the challenges that the development of quantum processors must face.

\subsection{Universal vs. non-universal gate sets and quantum advantage}

Our results also show a substantial difference between Clifford and Clifford $+$ \texttt{T} gate sets. While in the latter the violation degrees are distributed across the entire range of possible values, in the former, these are highly concentrated in specific points, with many states having the same maximal value of Svetlichny violation.

Also, it is clear that using circuits based on the Clifford set one can prepare states with a maximum violation value of the inequality \eqref{e:SVL_N_Qubits} for arbitrary $N$ because, for example, the GHZ state can be prepared with such gates. On the other hand, it is known that the Clifford set is non-universal, and that quantum algorithms built only out of Clifford gates plus preparations and measurements in the computational basis are efficiently simulable by a classical computer~\cite{Gottesman_1998}. This is one of the reasons why non-locality on its own cannot be the only resource needed for quantum advantage. There must be something else that allows us to distinguish between circuits that can be efficiently simulated with classical computers and those that cannot.

Our results indicate that the richness of the set of states produced with Clifford gates, quantified using genuine multipartite non-locality, is not as extensive as that of a universal gates set. The fact that Clifford circuits can be efficiently simulated with a classical computer, together with the differences between the corresponding non-locality histograms presented in this work (Figs.~\ref{f:Histograms_condensed} and~\ref{f:Clifford_Svetlichny}), could be used as a clue to understand the reasons for quantum advantage. We will study this problem with more detail in future works by analyzing other non-universal gate sets.

\subsection{\label{s:Conclusions}Open problems and future work}

Another relevant result is that the (relative) volume of the set of states violating the Svetlichny inequality given by Eq.~\eqref{e:SVL_N_Qubits} tends to decrease substantially as the number of qubits increases. In fact, it becomes almost zero already for five qubits (see Fig.~\ref{f:S_and_M_Violations}). This is in contrast with the asymptotic behavior of the volume of the set of entangled states. When using the entanglement entropy, the volume of the set of separable states tends to zero as $N$ increases~\cite{Entropy_as_N_grows,Hayden2006} (see also~\cite{Rosier2020}). Also, we have found that the violation fraction for Mermin's inequality is still non-negligible for five qubits. It is important to keep in mind the fact that even if a given state does not violate Svetlichny's inequality, it is not granted that there will not exist a stronger inequality revealing genuine multipartite non-locality for that state. Here we have focused on the inequality \ref{e:SVL_N_Qubits}. Out of our results, it is not possible to know what is the volume of the set of states displaying genuine multipartite nonlocality. But still, our findings open the door to keep inquiring on this subtle topic, indicating a complex behavior of the geometry of the set of quantum states as the number of qubits increases \cite{Makuta2025}.

Notice also that the number of correlation terms in Eq.~\eqref{e:SVL_N_Qubits} grows exponentially with the number of qubits. Thus, testing it on extant quantum processors is difficult. There are of course methods for addressing the challenge of scaling nonlocality inequalities on QPUs (see for example~\cite{Cabello-Non-locality-benchmarking}), but this is not our focus in this work. What is relevant for us is that simulating the violation of Eq.~\eqref{e:SVL_N_Qubits} is challenging even for a small number of qubits on today’s quantum processors. Comparing theory with experiment for $N<10$ is already very informative about the performance of a given QPU, as our results show.
    
By examining how non-locality is distributed in QRC for different gate sets, we have also gained insights into the impact of noise on near-term quantum devices. We also analyzed the dependence of the results obtained with regard to circuit depth and compared them with different measures of entanglement and quantum magic. Regarding entanglement measures, our results are consistent with previous efforts, extending the scope of the analysis to states which are uniformly distributed along the quantum state space. Globally, our results go beyond previous works in characterizing the connections between entanglement and non-locality.
     
Our findings could serve as the basis for developing calibration protocols in future works, as they provide valuable information on the quantum computer's ability to produce relevant resources ---such as genuine multipartite non-locality.
\vspace{-0.5cm}

\section*{Acknowledgements}
\footnotesize
Roberto Giuntini and Giuseppe Sergioli are partially supported by the
projects:

\begin{itemize}
    
    \item “CORTEX The Cost of Reasoning: Theory and Experiments”, funded by the Ministry of University and Research (Prin 2022, cod. 2022ZLLR3T).
    
    \item “Quantum Models for Logic, Computation and Natural Processes (Qm4Np)” funded by the Ministry of University and Research (Prin-Pnrr 2022 cod. P2022A52CR). 

\end{itemize}

Roberto Giuntini is partially funded by the T\"UV S\"UD Foundation, the Federal Ministry of Education and Research (BMBF) and the Free State of Bavaria under the Excellence Strategy of the Federal Government and the L\"ander, as well as by the Technical University of Munich-Institute for Advanced Study.

\section*{\label{s:CODE_AVAILABILITY}DATA AVAILABILITY}

The data that support the findings of this article are openly available~\cite{QuantumUnica_non_locality}.

\appendix

\section{\label{s:MerminAndSvetlichnyExplained}MERMIN AND SVETLICHNY INEQUALITIES}

We start by analyzing a simple two-qubit system and the violation of CHSH inequalities~\cite{CHSH_inequalities}. For the state associated with each generated QRC, we compute its maximal degree of violation. We consider from now on an experimental situation in which two dichotomic observables $A_{1}$, $A_{1}'$, $A_{2}$ and $A_{2}'$, with outputs  $1$ and $-1$ each, can be performed on each system. Let 
\begin{equation}
    S_{2}  = E\left( A_{1}A_{2} \right)  -  E\left( A_{1}A_{2}' \right)  + E\left( A_{1}'A_{2} \right)  +  E\left( A_{1}'A_{2}' \right), 
\end{equation}
\noindent where $E(A_{1}A_{2})$, for example, is the correlation function between $A_{1}$ and $A_{2}$, and so on with the primed variables. Based on the so-called local realism (or local hidden variable model), it is possible to obtain the CHSH inequality~\cite{CHSH_inequalities}:

\begin{equation}
    |S_{2}| \leq 2.\label{CHSH} 
\end{equation}  

Delving into the realm of quantum mechanics, our focus on dichotomic measurements allows us to confine our exploration to two-dimensional systems. In this case, the operator inside the modulus of the left-hand side of Eq.~\eqref{CHSH} becomes:

\begin{equation}
\begin{split}
O_{2} &= \vec{\sigma}\cdot \vec{a}_{0}\otimes\vec{\sigma}\cdot \vec{b}_{0} - \vec{\sigma}\cdot \vec{a}_{1}\otimes\vec{\sigma}\cdot \vec{b}_{0}+\vec{\sigma}\cdot \vec{a}_{0}\otimes\vec{\sigma}\cdot \vec{b}_{1} \\
& + \vec{\sigma}\cdot \vec{a}_{1}\otimes\vec{\sigma}\cdot \vec{b}_{1},
\end{split}
\end{equation}

\noindent where  $\vec{\sigma}=(\sigma_{x},\sigma_{y},\sigma_{z})$ with $\sigma_{x}$, $\sigma_{y}$ and $\sigma_{z}$ that are the Pauli operators and $\vec{a}_{0}$, $\vec{a}_{1}$, $\vec{b}_{0}$ and $\vec{b}_{1}$, are unit vectors representing the orientations of the polarizers of each pair. To each direction, there is a pair of angles associated. For a given state $\rho$ we maximize the quantity

\begin{equation}
V_{2}(\rho) = |\mbox{Tr}(\rho O_{2})|,
\end{equation}

\noindent among possible values of the angles. If the local realism is violated and consequently non-locality is present in a given state $\rho$ with regard to the chosen family of observables \footnote{Notice that, in principle, there could exist a more general family of observables for which a non-locality inequality is violated. We are restricting here the analysis to the CHSH-type observables (with variable angles).}, there should exist a combination of angles for which $S_{2}(\rho)=V_2(\rho)>2$. 

Now we turn to the study of systems with $2$, $3$, $4$ and $5$ qubits. When more than two particles are present, there exist different types of non-local states. First, one could start asking the question of whether there exists a local hidden-variable model reproducing the correlations of the system under study. This problem was addressed by Mermin~\cite{Mermin_1990} (see also~\cite{Alsina_2016} and~\cite{Werner_2001}). By requiring the condition of the local hidden variable model, it is possible to derive a set of inequalities based on the so-called Mermin polynomials $M_{N}$ (see for example~\cite{Alsina_2016}, section II-C). These polynomials are recursively defined in the following way: 

\begin{equation}
    M_{2}=\frac{1}{2}\left ( A_{1}A_{2}+ A_{1}'A_{2}+ A_{1} A_{2}' -A_{1}'A_{2}' \right ),
\end{equation}

and

\begin{align}
    M_{N}=\frac{1}{2}M_{N-1}\left ( A_{N}+A_{N}' \right )+\frac{1}{2}{M}_{N-1}'\left ( 
A_{N}- A_{N}'\right ),
\end{align}

\noindent where $M'_{N-1}$ is obtained from $M_{N-1}$ by exchanging all the zero-indexed  and one-indexed $A$'s. The local realism limit is given by~\cite{Alsina_2016}:

\begin{equation}
 |\langle M_{N} \rangle| \leq 1.\label{e:Mermin}
\end{equation}

For $N=2$ we recover one of the equivalent ways of writing the CHSH inequalities up to factor $1/2$. The violation of the $N$-particle Mermin inequality implies that the correlations cannot be modeled using local realism. However, in a multi-partite system, restricting the analysis to the simple absence of local hidden-variable models might yield a narrow perspective on the possible correlations involved. For example, one could have a state of three particles in which the first two are maximally correlated, while there is no correlation with regard to the third one, as is the case for a quantum system of three particles prepared in the state

\begin{equation}
|\psi\rangle = \frac{1}{\sqrt{2}}\left (|00\rangle+ |11\rangle  \right )|0\rangle.\label{e:partial_entanglement}
\end{equation}

The above state clearly displays non-local correlations between the first and second particles, but these correlations do not involve all the parties. In order to distinguish among the different types of correlations, G. Svetlichny and collaborators~\cite{Svetlichny_PRD, Seevinck_Svetlichny_PRL} developed a set of inequalities which, if violated, will reveal the presence of genuine multipartite non-locality (See also the discussion presented in Ref.~\cite{Collins_2002}). The Svetlichny inequalities have also been experimentally confirmed (see for example~\cite{Lavoie_2009}).

Consider the correlation functions $E(A_{1}^{(i)}A_{2}^{(j)}A_{3}^{(k)})$, which represent the expected value of the product of the measurement outcomes of single particle random variables $A_{1}^{(i)}$, $A_{2}^{(j)}$ and $A_{3}^{(k)}$ ($i,j,k$ are primed or not primed variables). Define

\begin{widetext}
\begin{equation}
    S_{3}  = E\left(A_{1}A_{2}A_{3}\right)  +  E\left(A_{1}A_{2}A_{3}'\right)  + E\left(A_{1}A_{2}'A_{3} \right)   
    + E\left(A_{1}'A_{2}A_{3}\right)- E\left(A_{1}'A_{2}'A_{3}' \right) -  E\left(A_{1}'A_{2}'A_{3}\right) 
    -E\left(A_{1}'A_{2}A_{3}'\right)  -  E\left(A_{1}A_{2}'A_{3}'\right), \label{e:SvetlichnyA}
\end{equation}
\end{widetext}

\noindent then, the three-qubits Svetlichny inequality reads:

\begin{align}
|S_{3}| \leq 4.\label{e:SvetlichnyB}
\end{align}

If a state violates the above inequality, then, it presents genuine three-partite non-locality.

\noindent In a quantum setting, consider the emission of triplets of particles in a pure quantum state from a source, for which their state is possibly unknown. During each execution of the experiment, one of the two possible alternative measurements is performed: $A_{1}$ or $A_{1}'$ on the first particle, $A_{2}$ or $A_{2}'$ on the second particle, and $A_{3}$ or $A_{3}'$ on the third particle. Each operator is of the form $\vec{\sigma}\cdot\vec{a}_{i}$, $\vec{\sigma}\cdot \vec{b}_{i}$ and $\vec{\sigma}\cdot \vec{c}_{i}$, with $\vec{a}_{i}$, $\vec{b}_{i}$ and $\vec{c}_{i}$, different unit vectors (with $i=0,1$). For the quantum case, we have $E( A_{1}A_{2}A_{3})=\mbox{tr}(\rho A_{1}\otimes A_{2}\otimes A_{3})$, with $A_{1}$, $A_{2}$ and $A_{3}$ Hermitian operators. Define the three-qubits Svetlichny operator by:

\begin{equation}
\begin{aligned}
    O_{3} &=  A_{1}\otimes A_{2}\otimes A_{3} 
    + A_{1}\otimes A_{2}\otimes A_{3}' 
    + A_{1}\otimes A_{2}'\otimes A_{3}  \\
    & + A_{1}'\otimes A_{2}\otimes A_{3}  - A_{1}'\otimes A_{2}'\otimes A_{3}' - A_{1}'\otimes A_{2}'\otimes A_{3}  \\
    & - A_{1}'\otimes A_{2}\otimes A_{3}'
    - A_{1}\otimes A_{2}'\otimes A_{3}'.
\end{aligned}
\end{equation}

\noindent For a given three-qubits state $\rho$, we maximize the quantity

\begin{equation}
V_{3}(\rho) = |\mbox{Tr}(\rho O_{3})|,\label{e:Three_qubits_maximize}
\end{equation}

\noindent for all possible values of the polarizers. As in the two-qubits case, a violation of the above inequality implies that the state has genuine three-partite non-locality. Thus, given a state $\rho$, we look for the angles which maximize \eqref{e:Three_qubits_maximize} and use that maximal value to quantify the eventual non-locality present in the state. We refer to the value obtained as \textit{violation level}. A state with no non-locality with regard to the chosen set of observables will display a violation level below the classical limit of 4. As is well known~\cite{Svetlichny_PRD}, for the three-qubit case, the maximum value of violation that can be reached with a quantum state is $4\sqrt{2}$. Thus, it is the maximum violation level reachable using quantum states. 

For the case of an arbitrary number $N$ of qubit systems, the inequality is built using the quantity~\cite{Seevinck_Svetlichny_PRL}:

\begin{equation}
    S_{N}^{\pm}=\sum_{I}\nu _{t\left ( I \right )}^{\pm}A^{i_{1}}_{1}\cdots A^{i_{N}}_{N},\label{e:Svelichny_N_Qubits_operator}
\end{equation}

\noindent in the above inequality, $\nu_{t}^{\pm }=\left ( -1 \right )^{\frac{t\left ( t\pm 1\right )}{2}}$, with $I=\left ( i_{1}, i_{2},\cdots, i_{N} \right)$, with each $i_{j}$ indicating whether a prime appears or not in the corresponding variable. The number $t\left ( I \right )$ indicates the number of times primes appear in $I$, and $\nu_{t}^{\pm}$ is a sequence of signs. Two types of inequalities are derived, namely

\begin{equation}
    |S_{N}^{\pm}|\leq 2^{N-1}.\label{e:SVL_N_Qubits}
\end{equation}

In order to describe the numerical optimization method used through this section, let us focus on the quantity $V_{3}(\rho)$ (the other cases are computed in a similar way). Full details can be found in the code repository shown in the \ref{s:CODE_AVAILABILITY} Section.

\begin{itemize}
\item Use the \textit{Numpy} Python library to write $V_{3}(\rho)$, using the density operator of a generated circuit. Notice that an expression such as $V_{3}(\rho)$ depends on abstract angle parameters.
\item Maximize the quantity $V_{3}(\rho)$ using the function \textit{minimize} from the Python library \textit{SciPy}.
\end{itemize}

As a result, for a given state $\rho$ which is a matrix associated to a quantum
random circuit, after the application of the above steps, one obtains the maximal violation value,
together with the angles for the spin orientations of the local observables. These can be easily converted into one qubit quantum circuits implementing the local observables with the desired angles.

\section{\label{s:Appendix_entanglement_measures}ENTANGLEMENT AND MAGIC QUANTIFIERS}





\subsection*{Tangle}

The $3$-tangle (or residual tangle) coincides with the modulus of a so-called hyperdeterminant which was introduced by Cayley~\cite{cayley1893collected}

\begin{equation}
\label{eqn:Tau3}
    \tau_{3}=4\left | H \det\left ( t_{ijk} \right ) \right |,
\end{equation}

\noindent for our work, the hyperdeterminant $H \det\left ( t_{ijk} \right )$ is a polynomial of order four in the amplitudes $t_{ijk}$. It can be expressed by using the wave function coefficients $\left \{ \psi_{000}, \psi_{001},\cdots , \psi_{111} \right \}$~\cite{P_rez_Salinas_2020} as

\begin{equation*}
\begin{aligned} 
\tau_{3} &= 4\left | d_{1}-2d_{2}+4d_{3} \right |\\
d_{1} &=  \psi_{000}^{2}\psi_{111}^{2}+\psi_{001}^{2}\psi_{101}^{2}+\psi_{100}^{2}\psi_{011}^{2}\\
d_{2} &= \psi_{000}\psi_{111}\psi_{011}\psi_{100}+\psi_{000}\psi_{111}\psi_{101}\psi_{010}\\
&+\psi_{000}\psi_{111}\psi_{110}\psi_{001}+\psi_{011}\psi_{100}\psi_{101}\psi_{010}\\
&+\psi_{011}\psi_{100}\psi_{110}\psi_{001}+\psi_{101}\psi_{010}\psi_{110}\psi_{001}\\
d_{3} &= \psi_{000}\psi_{110}\psi_{101}\psi_{011}+\psi_{111}\psi_{001}\psi_{010}\psi_{100}.
\end{aligned}
\end{equation*}

\subsection*{Multipartite systems with bipartite entanglement}


As computing the von Neumann entropy can pose computational challenges due to the need to compute the eigenvalues of the reduced density operator, it is often preferable, particularly in the context of multipartite systems~\cite{MeyerWallach2002}, which can be expressed as an average of the bi-partite entanglement between each qubit and the rest of the system~\cite{brennen2003} 
\begin{equation}
    \label{eqn:Emerson}
    Q=2-\left ( \frac{2}{n_{q}} \right )\sum_{i=1}^{n_{q}}\mathrm{Tr}[\rho_{i}^{2}],
\end{equation} 
\noindent here, the $\rho_{i}=\mbox{Tr}_{i}|\psi\rangle \langle\psi|$ represent the reduced density matrix of the $i$-th qubit, obtained by tracing out the remaining $n_{q}-1$ qubits. The parameter $n_{q}$ denotes the total number of qubits. It follows that $0\leq Q\leq 1$. 
\newpage
\subsection*{\label{s:Magic}Quantum Magic}
We employ the stabilizer 2-Rényi entropy \cite{Leone_2022}, commonly referred to as the ``magic measure". Let $|\psi\rangle$ denote a quantum pure state. Given a specific operator $C$, repeated measurements allow for the estimation of the probability $P\left ( s\mid  C\right )=\left | \langle s| C|\psi\rangle \right |^{2}$. We define the vector $\vec{s}=\left ( \vec{s}_{1},\vec{s}_{2},\vec{s}_{3},\vec{s}_{4} \right )$ comprising four $n$-bit strings, with the binary sum denoted as $\left | \vec{s} \right |\equiv\boldsymbol{s}_{1}\oplus \boldsymbol{s}_{2} \oplus \boldsymbol{s}_{3} \oplus \boldsymbol{s}_{4}$. In this framework, the stabilizer $2-$Rényi entropy is characterized by:

\begin{equation}
    M_{2}\left ( |\psi\rangle  \right )=-\log \sum _{\vec{s}}\left ( -2 \right )^{-\left \| \vec{s} \right \|}Q\left ( \vec{s} \right )-\log d,
\end{equation}

\noindent where $Q\left ( \vec{s} \right )=E_{C}P\left ( \boldsymbol{s}_{1}| C\right )P\left ( \boldsymbol{s}_{2}| C\right )P\left ( \boldsymbol{s}_{3}| C\right )P\left ( \boldsymbol{s}_{4}| C\right )$ represents the expectation value over randomized measurements of the Clifford operator $C$, and $d\equiv 2^{n}$ denotes the dimension of the Hilbert space corresponding to  $n-$qubits.

\onecolumngrid  

\section{\label{s:Histograms}SUPPLEMENTARY VIOLATION HISTOGRAMS AND ENTANGLEMENT}

\begin{figure*}[ht!]
\centering
\includegraphics[width=\textwidth]{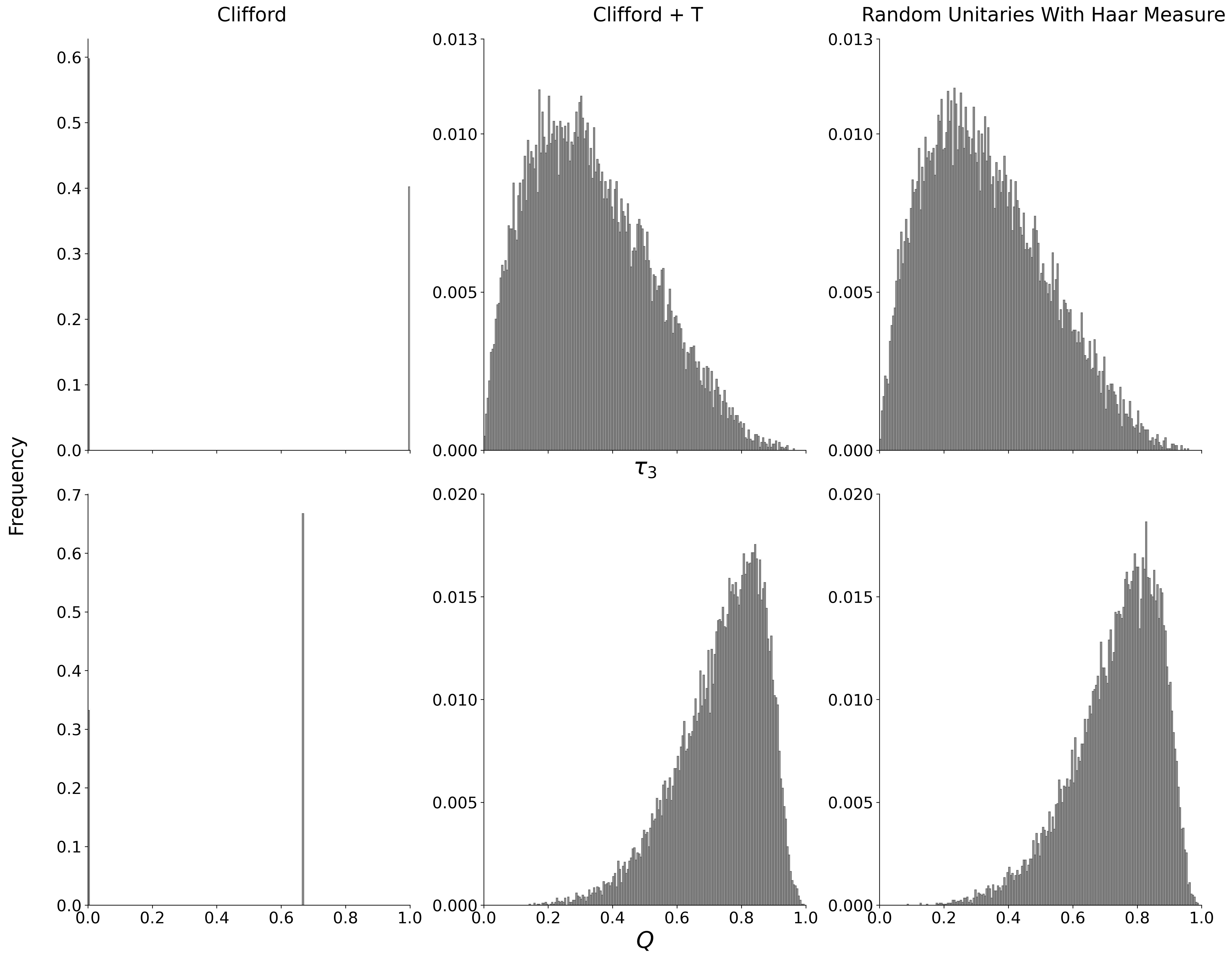}
\caption{\label{f:Distribution}Distributions of $\tau_{3}$ Eq. \eqref{eqn:Tau3} and $Q$ Eq. \eqref{eqn:Emerson} for $50{,}000$ three-qubit random circuits built using Clifford, Clifford $+$ \texttt{T}, and Haar-random unitaries, respectively.}
\end{figure*}

\FloatBarrier 
\begin{figure*}[ht!]
\centering
\includegraphics[width=0.9\textwidth]{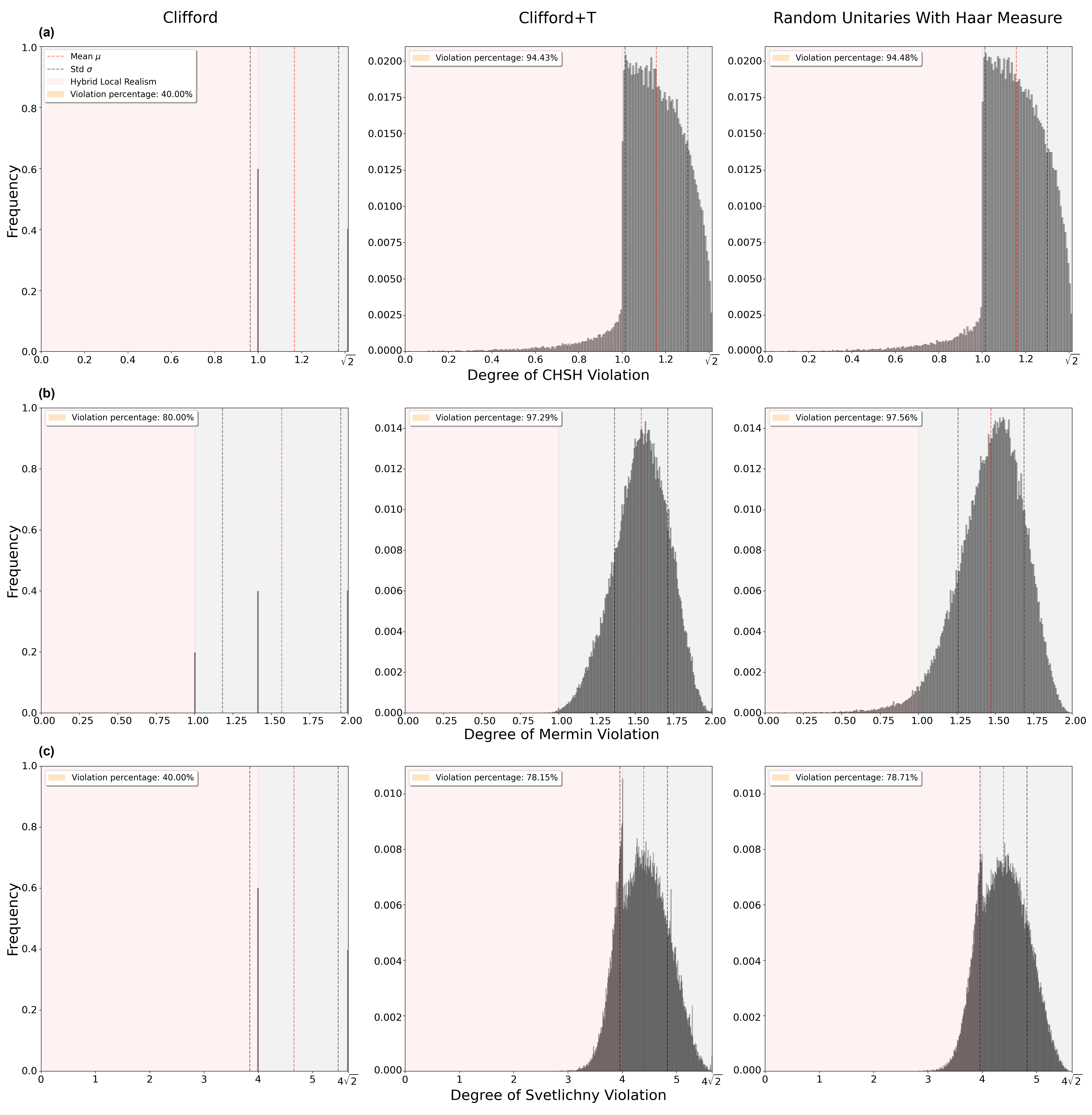}
\caption{\label{f:Histograms_Svetlichny}Violation levels of the (a) CHSH inequality (two qubits), (b) Mermin inequality and (c) Svetlichny inequality obtained with quantum random circuits. Columns show, from left to right: Clifford circuits, Clifford $+$ \texttt{T} (universal) circuits, and Haar distributed random unitaries. We have created $100{,}000$ instances of circuits for each case, and display the relative weights of the values obtained. The similarity between the histograms generated using Clifford $+$ \texttt{T} and Haar distributed random unitaries (which are produced using different and independent codes), serves to validate that our method for generating QRCs produces accurate results.}
\end{figure*}
\FloatBarrier

\FloatBarrier 
\begin{figure*}[ht!]
\centering
\includegraphics[width=0.8\linewidth]{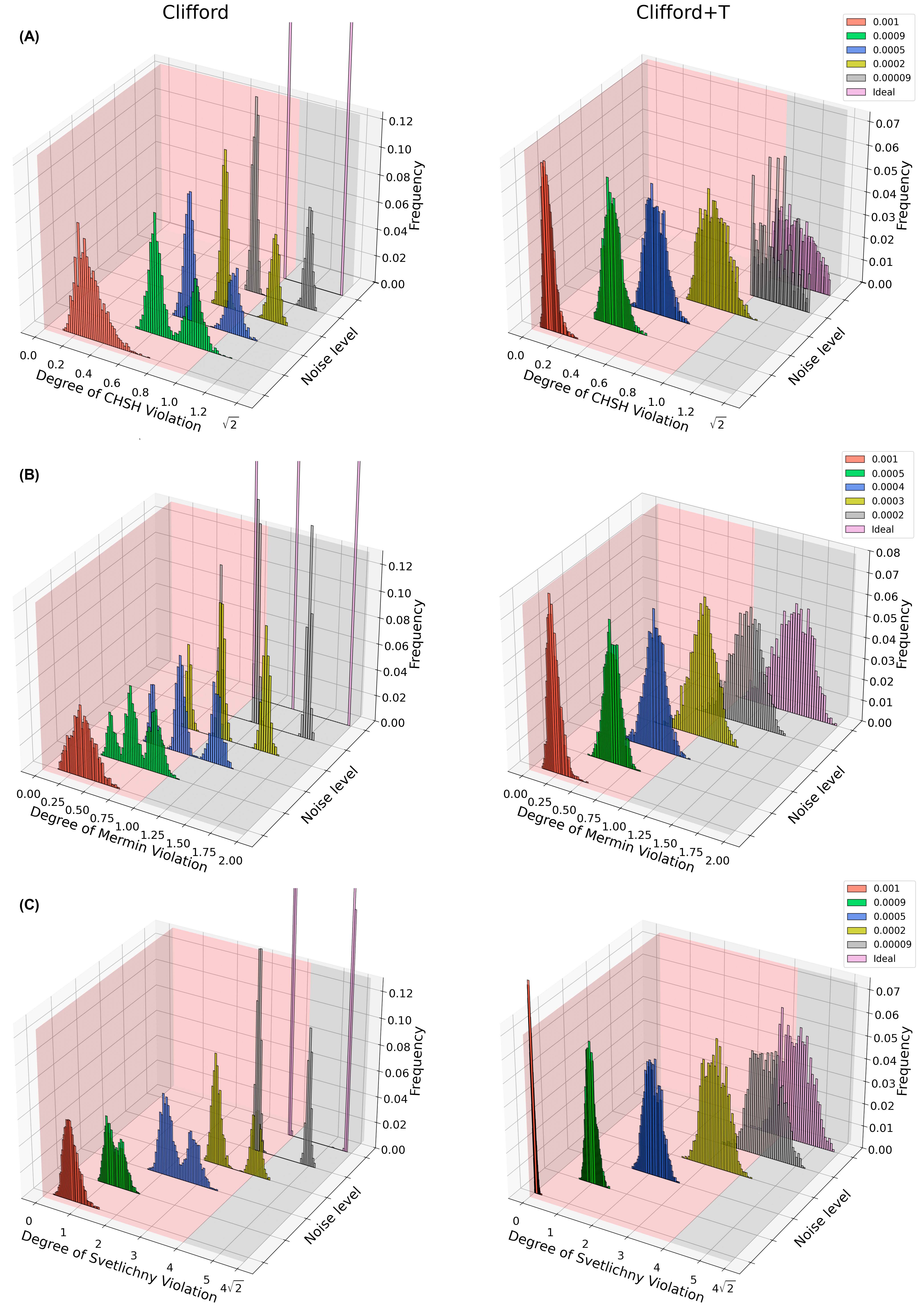}
\caption{\label{f:noise_behavior_all}Histograms of the violation levels of nonlocality inequalities for two- and three-qubit systems. 
For each panel, the values of the depolarizing-channel parameter are shown in the top-right corner. 
Each plot reports results from $3{,}000$ circuit instances for Clifford circuits (left column) and Clifford $+$ \texttt{T} circuits (right column). In (A) we show the two-qubits case using the CHSH inequality. For the three-qubit systems, we show the histograms for (B) Mermin inequality and (C) Svetlichny inequality.}
\end{figure*}
\FloatBarrier 

\FloatBarrier 
\begin{figure*}[ht!]
\centering
\includegraphics[width=\textwidth]{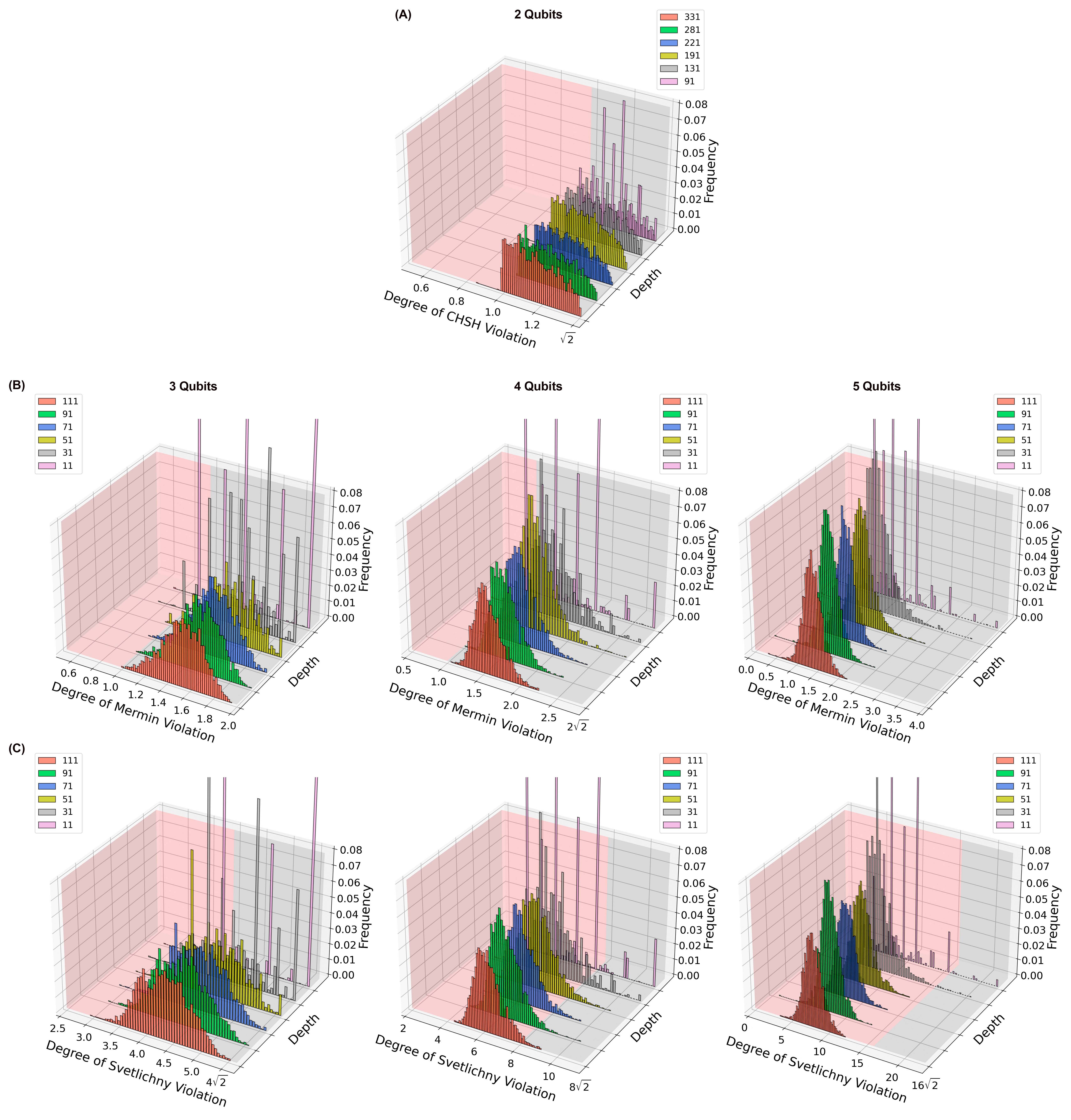}
\caption{\label{f:depth_variation_ct}Representation of noiseless violation level histograms for the Clifford $+$ \texttt{T} set as a function of the circuit depth. (A) CHSH inequality (two qubits), (B) Mermin inequality and (C) Svetlichny inequality from two to five qubits. The different depth values chosen are displayed at the top-right corner of each plot. Notice that for higher depth levels, the curves resemble those of states generated with random unitaries. But, in this case, the chance of obtaining a state with a near maximal violation level tends to decrease as the number of qubits grow. On the contrary, for lower depth levels, it becomes more likely to obtain states with a high violation degree.}
\end{figure*}
\FloatBarrier

\FloatBarrier 
\begin{figure*}[ht!]
\centering
\includegraphics[width=\textwidth]{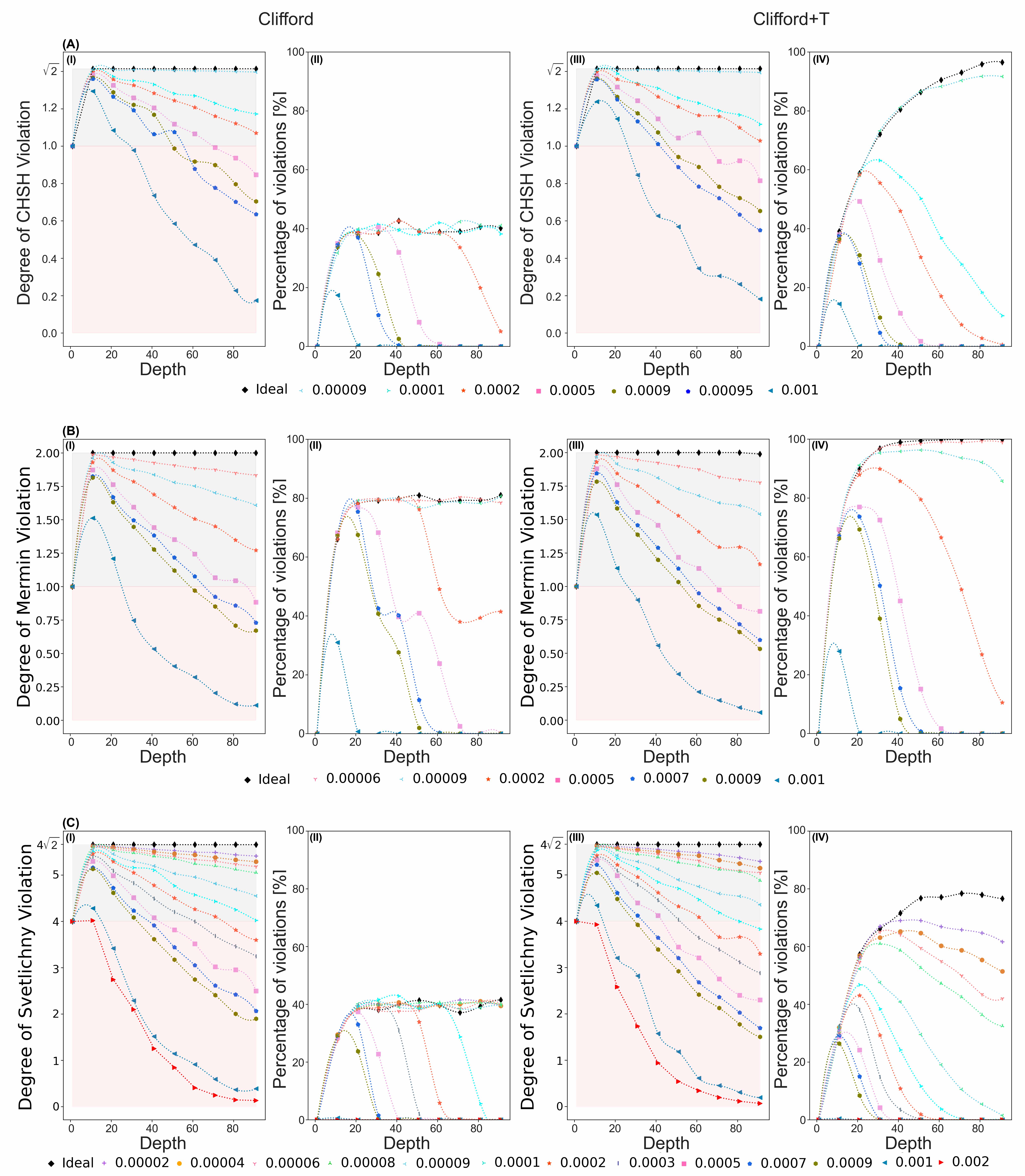}
\caption{\label{f:noise_and_depth_behavior_all}(A) CHSH inequality, (B) Mermin inequality and (C) Svetlichny inequality are examined, wherein figures (I) and (II) indicate the violation level and the fraction of violations as functions of circuit depth for Clifford gates, across varying noise levels. Figures (III) and (IV) present analogous results, albeit for the Clifford $+$ \texttt{T} gate set.}
\end{figure*}
\FloatBarrier

\FloatBarrier 
\begin{figure*}[ht!]
\centering
\includegraphics[width=\textwidth]{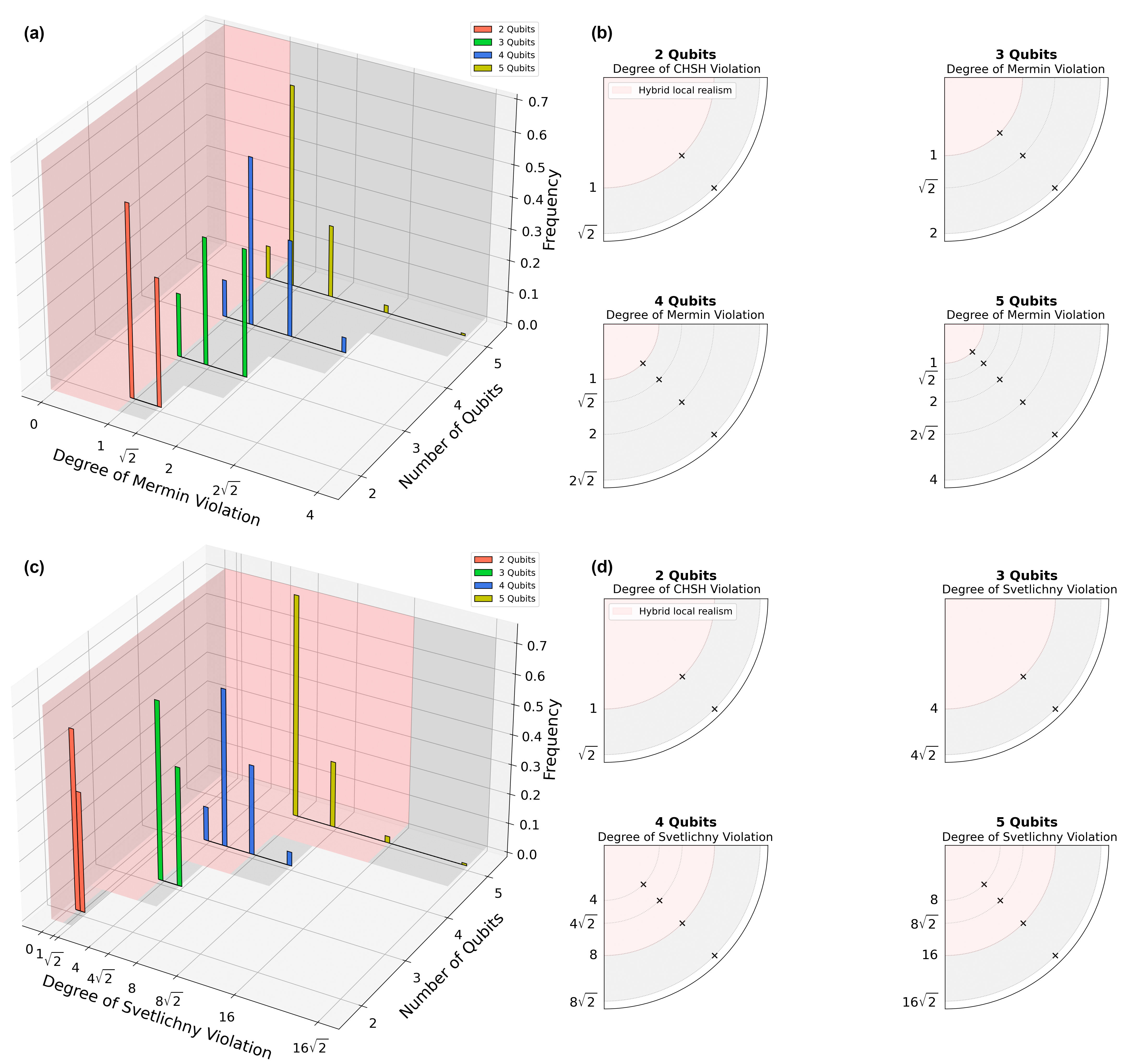}
\caption{\label{f:all_qubit_numbers_Sve}(a) Histograms of violation levels (without noise) of Mermin inequality for $100{,}000$ random circuits for $N = 3, 4, 5$ for states generated using the Clifford group. In (b) we plot the list of violation levels obtained for the Mermin inequality for $N = 3, 4, 5$. In (c) and (d) we do the same as in the previous case but for Svetlichny inequality. If all possible values were reached, then, we should observe a line segment. But, as remarked above, the figures obtained for the Clifford group have ``holes'', in the sense that the violation levels tend to concentrate around specific points. These plots should be compared with those of the histograms in Fig.~\ref{f:Histograms_Svetlichny} (and with the results presented in~\cite{monchietti2023}). For completeness, we have also included on (a), (b), (c), and (d) the histogram corresponding to the CHSH inequality. Using a suitable normalization, the CHSH inequality can be made a particular case of Mermin or Svetlichny inequalities for $N=2$. For the Mermin case, the right normalization implies that Tsirelson's bound becomes $\sqrt{2}$, while the LHV limit becomes 1.}
\end{figure*}
\FloatBarrier

\newpage
\twocolumngrid

\end{document}